\documentclass{aa}
\usepackage{graphicx}
\usepackage{txfonts}
\usepackage{subcaption}
\usepackage{lscape}            
\usepackage{placeins}         
\usepackage{booktabs}
\usepackage{float}   
\usepackage{subcaption}
\usepackage{placeins} 
\usepackage{afterpage}  
\usepackage{newtxtext,newtxmath}
\usepackage{float}
\usepackage{amsmath}    
\usepackage{graphicx}   
\usepackage[colorlinks=true,
            linkcolor=blue,
            citecolor=blue,
            urlcolor=blue]{hyperref}
\usepackage{natbib}
\bibpunct{(}{)}{;}{a}{}{,}

\begin{document}

   \title{The bar-driven central complex of NGC 6946: A panchromatic and spatially resolved view of its star formation history and interstellar medium}
\titlerunning{Panchromatic and spatially resolved view of NGC 6946}

  \author{
Santiago Álvarez-Córdoba\inst{1}
\and
Jacopo Fritz\inst{1}
\and
M. A. Lara-López\inst{2,3}
\and
Bernardo Cervantes Sodi\inst{1}
\and
L. S. Pilyugin\inst{4,5}
\and
Viviana Casasola\inst{6}
\and
Sami Dib\inst{7}
\and
Omar López-Cruz\inst{8}
\and
L. E. Garduño\inst{2,3}
\and
G. Valé\inst{2,3}
\and
I. A. Zinchenko\inst{9,5}
\and
M. E. De Rossi\inst{10,11}
\and
M.M. Puebla\inst{3}
\and
Vidhi Tailor\inst{6,12}
\and 
Gustavo Bruzual \inst{1}
\and
J. Zaragoza-Cardiel \inst{13}
\and
Margarita Rosado \inst{14}}

\authorrunning{Álvarez-Córdoba et al.}

\institute{
Instituto de Radioastronomía y Astrofísica, UNAM, Campus Morelia, A.P. 3-72, Morelia 58089, Mexico
\and
Instituto de Física de Partículas y del Cosmos, IPARCOS, Fac. C.C. Físicas, Universidad Complutense de Madrid, E-28040
Madrid, Spain
   \and 
   Departamento de Física de la Tierra y Astrofísica,  Fac. de C.C. Físicas, Universidad Complutense de Madrid (UCM), E-28040 Madrid, Spain
 \and
Institute of Theoretical Physics and Astronomy, Vilnius University, Sauletekio av. 3, 10257, Vilnius, Lithuania
\and
Main Astronomical Observatory, National Academy of Sciences of Ukraine, 27 Akademika Zabolotnoho St., 03143, Kyiv, Ukraine\label{MAO}
\and
INAF - Istituto di Radioastronomia, Via Gobetti
 101, 40129 Bologna, Italy
 \and
Max Planck Institute for Astronomy, K\"{o}nigstuhl 17, 69117, Heidelberg, Germany
\and
Instituto Nacional de Astrofísica, Óptica y Electrónica (INAOE), Luis E. Erro No. 1, Sta. Ma. Tonantzintla, Puebla, C.P. 72840, Mexico
\and
Astronomisches Rechen-Institut, Zentrum f\"{u}r Astronomie der Universit\"{a}t Heidelberg, M\"{o}nchhofstra{\ss}e 12-14, D-69120 Heidelberg, Germany\label{ARI}
\and
Universidad de Buenos Aires, Facultad de Ciencias Exactas y Naturales y Ciclo Básico Común, Buenos Aires, Argentina 
\and 
CONICET-Universidad de Buenos Aires, Instituto de Astronomía y Física del Espacio (IAFE), Buenos Aires, Argentina
\and
 Dipartimento di Fisica e Astronomia, Alma Mater Studiorum Università di Bologna, Via Piero Gobetti 93/2, I-40129, Bologna, Italy
 \and 
    Centro de Estudios de Física del Cosmos de Aragón (CEFCA), Plaza San Juan 1, 44001 Teruel, Spain
 \and
 Universidad Nacional Autónoma de México. Instituto de Astronomía. A.P. 70-264, 04510. Ciudad de México, México}
 
   \date{Accepted September 30, 2026}

  \abstract 
   {{We present a detailed multi-wavelength analysis of the nearby spiral galaxy NGC~6946, combining optical integral field spectroscopy (GCMS), infrared (\textit{Herschel}), and radio observations (THINGS and HERACLES).}} 
   {We aim to characterise both the integrated and spatially resolved properties of the stellar populations and interstellar medium (ISM), and to interpret them in the context of bar-driven galaxy evolution.}
   {We used the spectral energy distribution fitting code \texttt{SINOPSIS} to derive stellar population properties (stellar mass, star formation history), and different prescriptions to derive ISM properties including atomic and molecular gas masses and the dust mass, dust temperature, and dust emissivity index, combining optical, infrared, and radio datasets.}
{We find that the central $\sim 2.6$ kpc ($\sim 0.2 \,R_{25}$) hosts $\sim40\%$ of the stellar mass, as well as the bulk of the molecular gas and dust, and has sustained enhanced star formation over cosmic time. A major star formation episode occurred $\sim 0.6$ Gyr ago. The spatially resolved Kennicutt--Schmidt relation shows that molecular gas alone explains only part of the current SFR variance, while alternative prescriptions incorporating stellar mass and dynamical equilibrium pressure yield only modest improvements. We also find radial enhancements in the effective oxygen yield, $Y_{\mathrm{eff}}$, and gas mass fraction, $\mu$, relative to  surrounding regions, consistent with sustained gas accumulation. Together with the molecular gas streaming motions previously detected in CO and H$\alpha$ kinematic studies, our results are consistent with, and provide independent support for, the inward transport of gas driven by a system of nested bars.}
 {Our results are consistent with a scenario in which a recent episode of bar-driven gas inflow enhanced the central molecular gas reservoir and sustained elevated star formation, further supported by the elevated gas-to-dust ratio in the inner region, and the low global $M_{\mathrm{HI}}/M_{\mathrm{H}_2}$ ratio. These findings point to efficient HI-to-H$_2$ conversion, likely promoted by the combined effect of the secondary bar and the large dust content, ultimately regulating star formation. We also find indications consistent with a young stellar bar, given by the presence of a low-mass pseudo-bulge. The current decline in star formation despite large gas reservoirs suggests recent bar-driven quenching.}


   \maketitle
\nolinenumbers

\section{Introduction}
Galaxies, the building blocks of the Universe's large-scale structure, exhibit a wide range of physical properties reflecting their formation and evolution history. Since galactic emission spans the entire electromagnetic spectrum, panchromatic studies are required to capture the full range of physical phenomena and facilitate a comprehensive analysis. In the optical, the spectral energy distribution (SED) is mainly determined by the star formation history (SFH)  of the galaxy: this shapes the UV emission via young, massive stars, and the optical and near-infrared (NIR) via intermediate and low-mass stars. Metallicity also plays a role, affecting both absorption and emission lines strengths as well as colours. On the other hand, interstellar dust is responsible for reddening by efficiently absorbing UV and optical light, and re-emitting it in the infrared \citep{Draine2003}. As such, dust plays a crucial role at both short and long wavelengths. Gas also shapes the optical spectra  through emission lines from ionised elements and contributes at longer wavelengths through free-free radiation. This allows us to use SEDs to infer the physical processes and properties that  drive the evolution of galaxies.\\
To extract the optical SED, integral field spectroscopy (IFS) is an observational technique that is based on the construction of 3D spectral maps (two spatial dimensions, one spectral) using an integral field unit (IFU) that takes multiple simultaneous spectra across the field of view. Major IFS galaxy surveys include: the Calar Alto Legacy Integral Field Area
survey CALIFA \citep{CALIFA2012}, with $\sim$600 local galaxies ($0.005<z<0.03$) covering a significant luminosity range; the Mapping Nearby Galaxies at APO MaNGA \citep{MaNGA2015}, surveying $\sim$10000 nearby galaxies, sampling galaxies across $\sim$3 orders of magnitude in stellar mass ($M_*>10^9M_\odot$) without morphological or environmental restrictions, achieving $1-2$ kpc spatial resolution to study internal structures. Other surveys include the Sydney-AAO Multi-object Integral field spectrograph (SAMI) survey \citep{SAMI2015}, the ATLAS$^{\rm 3D}$ survey \citep{ATLAS2011}, the PPAK IFS Nearby Galaxies Survey (PINGS) \citep{Rosales2010}, and the Physics at High Angular resolution in Nearby GalaxieS with MUSE (PHANGS--MUSE) survey \citep{PHANGS2022} and Metal-THINGS \citep{MetalTHINGS2021}. \\
To complement the optical data and trace  the interstellar medium (ISM) components, the dust properties are also analysed. While dust constitutes only $\sim 1\%$ of the total ISM mass (with gas making up the remaining $\sim 99\%$, \citealt{Draine2011}), it is a critical component. Dust influences the energetic balance that regulates ISM heating and cooling \citep{Galliano2018} and drives several key processes. By attenuating and reddening starlight and nebular emission, dust grains introduce important uncertainties into measurements of galactic stellar masses, SFRs, metallicities, and the total energy budget of galaxies. Most relevant to this work, the formation of molecules (notably H$_2$) occurs on the surfaces of dust grains \citep{Hollenbach1971}, which subsequently shield these molecules from dissociating radiation \citep{Wakelam2017}. Thus, dust  facilitates the formation of molecules and the eventual formation of stars \citep[e.g.][]{Scoville2013}. The \textit{Herschel} Space Observatory, operating across the far-infrared (70–500 $\mu$m), samples the peak of the dust SED. This makes it sensitive to both the diffuse, cold dust ($T < 25$ K)  that dominates the total dust mass in galaxies \citep{Draine2007}, and the warmer dust ($T > 30$ K) radiating at shorter wavelengths that often dominates the total infrared luminosity.

On the other hand, as the bulk of the ISM mass is in the form of gas -- both atomic and molecular -- observations of the 21 cm line of atomic neutral hydrogen (HI) are fundamental for understanding the processes governing star formation, the dynamics and structure of the ISM, and the distribution of (dark) matter \citep{Walter2008, deBlok, Dib2021}. Atomic gas, primarily HI, dominates the mass budget of gas of most star-forming galaxies and is typically their most extended component, often stretching far beyond the stellar and molecular discs. The 21 cm emission line provides a powerful tool for mapping the distribution and kinematics of this phase of the ISM \citep{Thomas2004, Bigiel2012}. HI plays an important role as the indirect fuel for star formation \citep{Schaye2004, Bigiel2008}. Although stars form directly from molecular clouds, atomic hydrogen is the essential reservoir from which molecular gas forms. Consequently, the HI content directly influences a galaxy's evolutionary timeline and all properties linked to its star formation activity. Given that all galactic components are intrinsically linked, multi-wavelength studies are essential to uncover the relationships between them and establish the causal chains driving galactic evolution.

At the core of our study is the influence of stellar oval distortions in the evolution of galaxies. It is well established that stellar bars play a pivotal role in the secular evolution of their host discs. They act as efficient drivers for the redistribution of angular momentum within galaxies, facilitating a net mass transfer from the outer disc towards the circumnuclear regions \citep{Weinberg1985,Shlosman1989}. As gas is funnelled inwards by the bar potential, it accumulates and eventually forms a nuclear disc or pseudo-bulge. The formation epoch of this nuclear structure can be used to date the onset of the bar formation. Theoretical studies indicate that such nuclear discs form relatively quickly after bar formation, on timescales of $\sim 10^8$ years \citep{Athanassoula1992b, Athanassoula1992, Kormendy2010}.\\
Several studies have investigated the role of stellar bars in driving large molecular gas concentrations towards galactic centres, leading to enhanced star formation, increased dust content, and central rejuvenation \citep{Garcia2003,Casasola2011,Chown2019, Yu2022, Sormani2023, SanchezOs2026,Sanchez2026}. Bars also influence global galaxy properties, as evidenced by colour diagnostics: strong-barred galaxies tend to occupy the red region of colour–magnitude and colour–colour diagrams and exhibit a significant excess of high-metallicity values, while unbarred and weakly barred systems extend towards bluer colours \citep{Lee2012}. This suggests an evolutionary sequence linking different morphological types \citep{Vera2016}. These effects may be even more pronounced in nested bar systems -- such as NGC~6946 -- than in single-bar galaxies, as suggested by the results of \citet{Li2023}.

An important aspect of this work is the role of high central mass concentrations, which, according to several works, can weaken and/or destroy bars \citep[e.g.][]{Roberts1979, Norman1996, Sellwood1999, Athanassoula2005}. This implies that some currently unbarred spirals may have hosted bars in the past, and that bars can be transient and recurrent over a galaxy's lifetime. Consequently, bars in present-day disc galaxies may have formed at different epochs under varying conditions \citep{Bournaud2002, Kormendy2004,  Gadotti2006}.\\
In this context, NGC 6946, the object we study in this paper, is a spiral galaxy also known as the `Fireworks Galaxy' due to the large number of observed supernovae (SNe). Over the last $\sim 100$ years, 11 core-collapse events have been observed, making NGC 6946 an extreme case of SN production \citep{Eldridge2019}.  The morphology of NGC 6946 is somewhat complicated. The classification assigned to this galaxy according to \cite{Vaucoleurs1991} is SAB(rs)cd,  indicating the presence of a weak or intermediate bar (not dominant), with indications of a partial or distorted stellar and gas ring. The  parameters for NGC 6946 that will be used throughout this work are listed in Table \ref{tab:parametros}.
Dynamically, NGC 6946 exhibits a high degree of complexity, which is readily observable due to its relative proximity. This complexity is characterised by the presence of at least two nested stellar bars \citep{Regan1995, Elmegreen1998, Schinnerer2007}. Given its distinct nature compared to the rest of the disc, the central region of NGC~6946 is the main focus of our discussion in Section~\ref{sec:discussion}. There, we interpret its stellar population and ISM properties in the context of the galaxy's stellar bars. For example,  \cite{Romeo2015} analysed the structure and stability of its molecular disc, identifying a double-disc structure (nuclear + main) with a clear transition at a radius of $R \sim 1$ kpc from the galactic centre, where the dynamics of the molecular gas change. 
In addition, NGC 6946 lacks a classical bulge but hosts a small pseudo-bulge. Furthermore, there is no evidence of an active galactic nucleus (AGN) that could otherwise dominate or contaminate the central infrared emission \citep{Kormendy2010}, suggesting that its central structure is indeed a secular feature built by bar-driven processes.
In light of this, we perform a multi-wavelength analysis of the global and spatially resolved properties of NGC 6946. We aim to correlate the features of the different components of the ISM with the stellar populations to identify the causes and consequences of the observed properties, especially in the central ($\sim 2$ kpc) region of the galaxy. This region is distinct from the rest of the disc, and we discuss whether this difference is due to the complex dynamics and stellar bar structures observed in NGC~6946. The primary dataset for this work comes from the Metal-THINGS survey \citep{MetalTHINGS2021, MetalTHINGS}, which provides new optical IFU observations.  This paper is organised as follows. Section~\ref{sec:data} presents the optical, infrared, and radio datasets used in this work. Section~\ref{sec:methods} details the methodology, including SED fitting  and the derivation of various ISM properties. Section~\ref{sec:results} presents our main results, including the spatially resolved and integrated properties of NGC~6946. Section~\ref{sec: rings} describes a tilted-ring analysis of the galactic disc while section~\ref{sec:KS} provides an analysis of the Kennicutt-Schmidt (KS) relation in the galaxy. In section \ref{sec:discussion} we present the discussion of our results, and  section~\ref{sec:conclusions} summarises the main conclusions.

\begin{table}
    \centering
    \caption{Parameters of NGC 6946 adopted in this study.}
    \label{tab:parametros}
    \begin{tabular}{lc}
        \hline
        \textbf{Parameter} & \textbf{Value} \\
        \hline
        Distance$^a$ & $7.83 \pm 0.29$ Mpc \\
        Inclination$^b$ $\langle i\rangle$ & $32.8^\circ$ \\
        P.A.$^b$ & $242.7^\circ$ \\
        Centre Position$^c$ &
        $\alpha = 20^{\text{h}}34^{\text{m}}52.2^{\text{s}}$, 
        $\delta = +60^\circ 09'14.4''$ \\
        $R_{25}$$^d$ & $343\arcsec\,(13.02\,\mathrm{kpc})$ \\
        $R_{\mathrm{eff,NUV}}$$^e$ & $203.37\arcsec\,(7.72\,\mathrm{kpc})$ \\
        \hline
    \end{tabular}
    \tablebib{
        $^{(a)}$~\citet{Murphy2018};
        $^{(b)}$~\citet{deBlok};
        $^{(c)}$~\citet{Trachternach};
        $^{(d)}$~\citet{Karachentsev2004};
        $^{(e)}$~Valerdi et al. (in prep.).
    }
\end{table}

\section{Data}
\label{sec:data}

\subsection{Optical IFS data: Metal-THINGS}
\label{sec: Metal-THINGS}

The optical data used in this paper originate from the most recent data release of the Metal-THINGS survey \citep{MetalTHINGS2021}, which provides IFS of many nearby galaxies. This survey complements the neutral atomic gas (21 cm) observations from THINGS \citep{Walter2008} and infrared data from the SINGS \citep{SINGS} and KINGFISH \citep{Kingfish2011} surveys, with spatially resolved optical spectroscopy. Metal-THINGS aims to enable comprehensive multi-wavelength analysis of ionised gas, neutral gas, dust, and stellar populations to understand galaxy evolution processes, including feedback mechanisms, interactions, and accretion events.\\
Observations were conducted using the 2.7 m Harlan Schmidt telescope at McDonald Observatory with the George and
Cynthia Mitchell Spectrograph (GCMS) IFU, formerly known as VIRUS-P \citep{Hill2008}. The instrument provides a $100\times102$ arcsecs field of view per pointing, enabling coverage of nearby galaxies with multiple pointings. The IFU contains 246 fibres arranged in a fixed pattern, each covering a 4.2 arcsec diameter in the sky. Given the $\sim30\%$ filling factor, three dither positions were employed to achieve $90\%$ spatial coverage \citep{MetalTHINGS2021}. NGC 6946 was observed, for a total of 12 pointings, in the red configuration, covering $4400-7000$~\AA at FWHM = 5.4~\AA~resolution \citep{MetalTHINGS}. At the distance of NGC 6946, each fibre corresponds to a $\sim158$ pc physical scale. Similar to \cite{Garduno2023}, the fibres are transformed to a tridimensional cube convolving them by a Gaussian of FWHM = 4.16 arcsec keeping a pixel size of 4.16 arcsec.

\subsection{Infrared photometric data: Herschel}
To expand our study of NGC 6946, we incorporated far-infrared (FIR) data from the \textit{Herschel Space Observatory} \citep{Herschel2010A}, equipped with a 3.5-meter diameter mirror and operated across a wavelength range of 55 to 672 $\mu$m. 
For this work, we use data from the two photometric instruments: The Photodetector Array Camera and Spectrometer PACS \citep{Poglitsch2010}  and the Spectral and Photometric Imaging Receiver SPIRE \citep{Griffin2010}. The galaxy was observed by PACS at 100 and 160 $\mu$m with spatial resolutions of $6.7\arcsec \times 6.9\arcsec$ and $10.7\arcsec \times 12.1\arcsec$, respectively, and at the default SPIRE wavelengths 250, 350, and 500 $\mu$m with resolutions of $18.7\arcsec \times 17.5\arcsec$, $26.3\arcsec \times 23.4\arcsec$, and $38.1\arcsec \times 35.1\arcsec$, respectively (OBSIDs: 1342191947, 1342191948, 1342188786).

\subsection{Radio centimetre and millimetre data: THINGS and HERACLES}
We used 21 cm line emission data tracing atomic neutral hydrogen from The HI Nearby Galaxy Survey (THINGS, \citealt{Walter2008}). This high-resolution survey ($\sim6''$) of HI emission in 34 nearby galaxies was obtained using the Very Large Array (VLA), whose scientific objectives focused on ISM characteristics related to galaxy morphology, star formation, and mass distribution across the Hubble sequence. The THINGS science goals required the aforementioned high angular resolution to resolve neutral atomic gas complexes, trace HI fine structure, and resolve spiral arms. We employed the integrated HI map (moment 0) for NGC 6946 from the robust weighting reduction, with final beam parameters: B$_{\text{maj}} = 4.93''$, B$_{\text{min}} = 4.51''$, BPA = $1.1^\circ$, and pixel size = $1.5''$. Detailed descriptions of sample selection, observational set up, and data reduction procedures are provided in \cite{Walter2008}.\\
To map the molecular gas distribution in NGC 6946, we utilised data from the Heterodyne Receiver Array CO Line Extragalactic Survey HERACLES \citep{Leroy2009}. This survey targets CO emission from 18 nearby galaxies that overlap with both the THINGS and SINGS surveys, enabling coordinated multi-wavelength analysis. HERACLES employed the HERA receiver array on the IRAM 30-m telescope\footnote{IRAM is supported by CNRS/INSU (France), the MPG (Germany) and the IGN (Spain).} to map the CO($J=2\rightarrow1$) line across the entire optical disc (defined by the isophotal radius $R_{25}$, 25\,mag\,arcsec$^{-2}$ in the $B$ band) of each target. The data achieve $13''$ angular resolution and 2.6 km s$^{-1}$ velocity resolution, sufficient to resolve giant molecular cloud (GMC) associations \footnote{Giant molecular cloud associations (GMAs) are large-scale concentrations of molecular gas with masses $\sim 10^7\,M_\odot$ and sizes from a few hundred parsecs to 1 kpc, about an order of magnitude larger than individual GMCs  \citep[e.g.][] {Lo1987, Sakamoto1999, Casasola2007}.} and trace the molecular gas distribution. For NGC 6946, the coverage extends slightly beyond the optical disc, enabling comprehensive comparisons between molecular gas, stellar mass, atomic gas, and dust distributions. Further details on data processing and reduction techniques are provided in \cite{Leroy2009}.

\section{ Methodology}

\label{sec:methods}
In this section, we describe the tools used to derive the physical properties from our multi-wavelength dataset. First, we describe the treatment of the IFU data to extract stellar and ionised gas kinematics, as well as stellar population properties. Subsequently, we outline the modelling of the FIR emission used to derive dust properties. Finally, we present the methodology for calculating gas masses from radio observations. For the IFU data, we applied a correction using the \cite{Cardelli1989} extinction curve, to account for the dust of the Milky Way in the LoS of NGC 6946. In addition, a mask was created for the stars in the field of view of NGC 6946, in order to perform a cleaner analysis without the interference of objects not belonging to the galaxy, using both the Aladin Sky Atlas database \citep{Aladin}  and our own optical spectra  to reinforce their identification.

\subsection{Stellar kinematics}
We employed the Penalized Pixel-Fitting code (\texttt{pPXF}, \citealt{CappellariCopin2003,Cappellari}) to extract spatially resolved stellar kinematic properties for NGC 6946. This spectral fitting method uses mathematical penalisation to stabilise solutions and regularise against unphysical results or noise-dominated fits. The \texttt{pPXF} algorithm  is designed mainly to derive stellar and gas kinematics. It achieves this by finding the optimal linear combination of stellar population synthesis templates that best fits the observed spectrum across the full wavelength range. We applied \texttt{pPXF} to extract stellar kinematics and infer redshift for each spatial element using \texttt{E-MILES} SPS templates \citep{EMILES}.  The spatially resolved stellar velocity map  was then fed to the spectral synthesis code \texttt{SINOPSIS} (see Section \ref{sec:ssp_features}).

\subsection{Stellar population features}
\label{sec:ssp_features}
In the following step, we used the SImulatiNg OPtical Spectra wIth Stellar populations models code \texttt{SINOPSIS} \citep{Fritz2007,Fritz2011,Fritz2017} to reproduce key observed spectral characteristics by combining theoretical simple stellar population (SSP) spectra of different ages, thereby enabling derivation of stellar mass, SFH, mean stellar ages, and dust attenuation. \texttt{SINOPSIS} operates by minimising differences  between observed and model spectra through comparison of continuum fluxes in multiple narrow bands and equivalent widths (EWs) of both emission and absorption lines. Model construction employs SSP spectra of different ages with a common, predetermined metallicity. The fitting procedure tests different metallicity sets to identify optimal values, though \cite{Fritz2007} note that satisfactory fits can be achieved across multiple metallicity values, indicating this approach does not yield unique metallicity solutions. When fitting observed spectra with a single metallicity value, the assumption is that this value corresponds to the dominant emitting SSP. The model spectrum is constructed by summing individual SSP spectra, each weighted by its stellar mass contribution and modified by age-dependent extinction following the selective extinction hypothesis \citep{Calzetti1994,Calzetti2000,CharlotnFall2000}. For this paper, the star formation rate (SFR) of SSP younger than $2\times 10^7$ years (i.e. in the first four age bins) was kept constant within this range, and we only used four different values of extinction for  the SSPs, resulting in a lower number of free parameters. Primary outputs include total stellar mass (\( M_* \)), time-resolved SFH, stellar age-dependent dust extinction, and mean stellar population ages. For stellar mass estimation, we considered  stellar and remnant mass, corresponding to mass within stellar objects, including nuclear-burning stars, white dwarfs, neutron stars, and stellar black holes, excluding mass ejected into the ISM, according to the definition of \cite{Longhetii2009}.

\subsection{SINOPSIS setup for NGC 6946}
\label{ConfSINOPSIS}

The SED fitting for NGC 6946 was performed with \texttt{SINOPSIS} using a set of basic parameters. The model was configured with a spectral resolution of 5.4\,\AA\ FWHM and normalised at 6650\,\AA. The SFH was reconstructed using non parametric prescription based on the Charlot \& Bruzual (in prep.) SSP models, allowing independent SFRs and extinction for different age bins. The fitting utilised five stellar metallicities (0.002, 0.008, 0.014, 0.03, 0.04), the \cite{Calzetti1994} extinction law and a \cite{Chabrier2003} Initial Mass Function (IMF).  A total of 15 SSPs were used, covering an age range from 0 to $\sim$14\,Gyr, providing a fine sampling of young and intermediate ages. For each population, \texttt{SINOPSIS} fits the mass formed at that age (controlled by the \texttt{SFR} parameter) and the extinction affecting that population (controlled by the colour excess \texttt{E(B-V)}). 

As with any spectral fitting technique, the recovered SFHs are subject to well-known degeneracies between stellar age, metallicity and dust attenuation. These degeneracies are further exacerbated by the limited wavelength coverage of the spectra (4600–7000 \AA), which provides a restricted set of diagnostic spectral features and consequently reduces the temporal resolution with which the SFH can be reconstructed. Consequently, the recovered SFHs should be interpreted in terms of broad evolutionary phases rather than discrete, precisely dated star formation episodes. Although the spectral fitting employs a basis of 15 SSP ages to provide sufficient flexibility in reproducing the observed spectra, the limited wavelength coverage and the resulting age–metallicity–dust degeneracies do not justify interpreting the individual SSP contributions as independent events. We therefore rebin the recovered SFHs into four broader age intervals, which provide a more robust representation of the underlying SFH and avoid overinterpreting the intrinsic temporal resolution of the data. Accordingly, features such as the enhancement in star formation around $\sim$0.6 Gyr ago should be regarded as evidence of elevated star formation over a period of several hundred million years, rather than as a sharply defined instantaneous burst.

\subsection{Ionised gas emission lines}

We used our own routine (Osorio-Quiroga et al. in prep.) to fit gas emission lines using Gaussian functions to model each line and measure the  redshift, intensity, and broadening of gas emission lines across the entire galaxy. To achieve this, we used the emission-only spectrum, obtained by subtracting the purely stellar spectrum derived from the \texttt{SINOPSIS} fit from the observed spectrum. This step accounts for potential absorption components originating from stellar atmospheres, which can be especially crucial for the H$\alpha$ and H$\beta$ lines. Subsequently, Gaussian curves were fitted to each of the spectral lines, with the central wavelength (redshift), line width (velocity dispersion) and area under the curve (integrated flux or intensity, with its respective uncertainty) as parameters, including H$\beta$, {[N~I]} $\lambda\lambda$5198,5200, 
{[O~III]} $\lambda$5007,         
{[O~I]} $\lambda$6300,           
{[N~II]} $\lambda$6548,         
H$\alpha$,             
{[N~II]} $\lambda$6583,        
{[S~II]} $\lambda$6716 and        
{[S~II]} $\lambda$6731        lines. For the quantities derived from the emission lines, we impose a S/N$\geq 2$.

The code initialises the fitting taking an initial value for the redshift from the stellar kinematic fit. The parameter fitting is performed via non-linear least squares minimisation using the \texttt{curve\_fit} method from the SciPy library. Once we derived the emission lines parameters, we corrected for internal extinction caused by the interstellar dust of NGC\,6946, using the theoretical ratio   for HII regions for Case B recombination: $  \text{H}\alpha/{\text{H}\beta} = 2.86$ \citep{Oster2006agna} and adopting the extinction curve normalised to the V band from \cite{Cardelli1989} with a ratio of total to selective extinction of \(R_V = A_V / E(B-V) = 3.1\).

\subsection{Dust properties}
\label{methods:dust}
Once the total fluxes and their uncertainties were determined for each FIR band, we performed a SED fitting using a modified black-body (MBB) model \citep{Smith2010}. The MBB equation is given by
\begin{equation}
    F_\nu(T) = \frac{M_d}{D^2} \kappa_0 \left(\frac{\nu}{\nu_0}\right)^{\beta} \frac{2h\nu^3}{c^2} \frac{1}{e^{h\nu/kT} - 1}
    \label{eq:mbb}
,\end{equation}

where $F_\nu(T)$ is the flux density at frequency $\nu$, $M_d$ is the dust mass, $D$ is the distance to the source, $\kappa_0$ is the dust absorption coefficient at reference frequency $\nu_0$, $\beta$ is the dust emissivity index, which characterises the slope of the low frequency tail of the MBB spectrum depends on the properties of the dust grains, $h$ is Planck's constant, $k$ is Boltzmann's constant, and $c$ is the speed of light. We adopted $\kappa_0 = 0.192~\text{m}^2~\text{kg}^{-1}$ at the reference wavelength of 350~$\mu$m \citep{Draine2003}. The fitting procedure simultaneously constrained the dust mass, $M_d$, emissivity index, $\beta$, and dust temperature, $T_d$, through $\chi^2$ minimisation.

Another approach to applying the MBB fit and achieving better-resolved results (see Section \ref{sec:dust}) is to fix the value of $\beta$. We set this parameter based on the distribution derived from the five-band SED fitting. The flux ratio, $F_{\nu_1} / F_{\nu_2}$, can be written as

\begin{equation}
    \frac{F_{\nu_1}}{F_{\nu_2}} = \left(\frac{\nu_1}{\nu_2}\right)^{3+\beta} \frac{e^{h\nu_2/kT} - 1}{e^{h\nu_1/kT} - 1}
    \label{eq:flux_ratio}
.\end{equation}

This approach allows for an independent estimation of $T_d$ and consequently of $M_d$ using equation \ref{eq:mbb}. To perform a consistent pixel-by-pixel analysis, all images were convolved and reprojected to a common resolution corresponding to the poorest resolution of the dataset: the SPIRE 500~$\mu$m band. This was achieved using a library of convolution kernels  designed for \textit{Herschel} instruments, which account for their distinct beam shapes \citep{Aniano2011}.\\
Flux calibration uncertainties, taken from the PACS and SPIRE observer's manuals\footnote{\url{http://herschel.esac.esa.int/Docs/PACS/pdf/pacs_om.pdf}}\footnote{\url{http://herschel.esac.esa.int/Docs/SPIRE/pdf/spire_om_v24.pdf}}, are 5\% and 7\%, respectively. Fluxes were measured in apertures encompassing the entire galaxy with standard background subtraction using source-free circular regions (e.g. \citealt{Ciesla2012}). Instrumental, correlated, uncorrelated, confusion, and large-scale uncertainties were evaluated and found to be negligible compared to calibration uncertainties. The resulting fluxes and total uncertainties for each Herschel band are reported in Table~\ref{tab:fluxes}. Our FIR fluxes agree with those reported in the DustPedia database \citep{Clark2018} within uncertainties, validating our results.

\begin{table}
    \centering
    \caption{Total background-corrected fluxes for NGC 6946.}
    \label{tab:fluxes}
    \begin{tabular}{lcc}
        \hline
        \textbf{Band} & \textbf{Flux (Jy)} & \textbf{Uncertainty (\%)} \\
        \hline
        PACS 70 $\mu$m & 249.013 $\pm$ 12.452 & 5.00 \\
        PACS 100 $\mu$m & 457.423 $\pm$ 22.872 & 5.00 \\
        PACS 160 $\mu$m & 509.187  $\pm$ 25.460 & 5.00 \\
        SPIRE 250 $\mu$m & 251.220  $\pm$ 17.645 & 7.02 \\
        SPIRE 350 $\mu$m & 98.778  $\pm$ 6.956 & 7.04 \\
        SPIRE 500 $\mu$m & 35.522 $\pm$ 2.539 & 7.15 \\
        \hline
    \end{tabular}
\end{table}

\subsection{Atomic and molecular gas properties}

To estimate the total (and spatially resolved) atomic gas mass,  we used the integrated HI map (moment 0) for NGC 6946 from the robust weighting calibration from the THINGS  survey and the prescription to calculate the HI mass, using equation (3)  from \cite{Walter2008}.
The uncertainties were estimated taking into account  the uncertainty in the flux, dominated by the flux calibration accuracy, reported as $\sim 5\%$ by \cite{Walter2008}, propagating this  along with the uncertainty in the distance, we generated spatially resolved maps of the HI mass for NGC 6946. 

Similarly,  we obtained the $^{12}$CO(2--1) emission line data from the HERACLES survey. We reproduced a spatially resolved map of the H$_2$ mass by adopting a constant CO-to-H$_2$ conversion factor of $X_{\text{CO}} = 2.0 \times 10^{20} \, \text{cm}^{-2} (\text{K km s}^{-1})^{-1}$ with a $\pm 30\%$ uncertainty \citep{Bolatto2013}. However, for our radial analysis, we also adopted and discuss different prescription for the $X_{\text{CO}}$ conversion factor, finding that the overall trend persists in all cases. We also assumed a $^{12}$CO line ratio of $R_{21} = I_{\text{CO(2--1)}} / I_{\text{CO(1--0)}} = 0.64 \pm 0.09$ from \cite{denBrok2021}, typical for nearby, massive, star-forming spiral galaxies like NGC 6946.

\section{Spatially resolved and integrated properties}
\label{sec:results}

\subsection{Stellar populations}
\subsubsection{Stellar mass and stellar age}
\label{ss:stellar mass and age}
\begin{figure}
    \centering
    \includegraphics[width=0.45\textwidth]{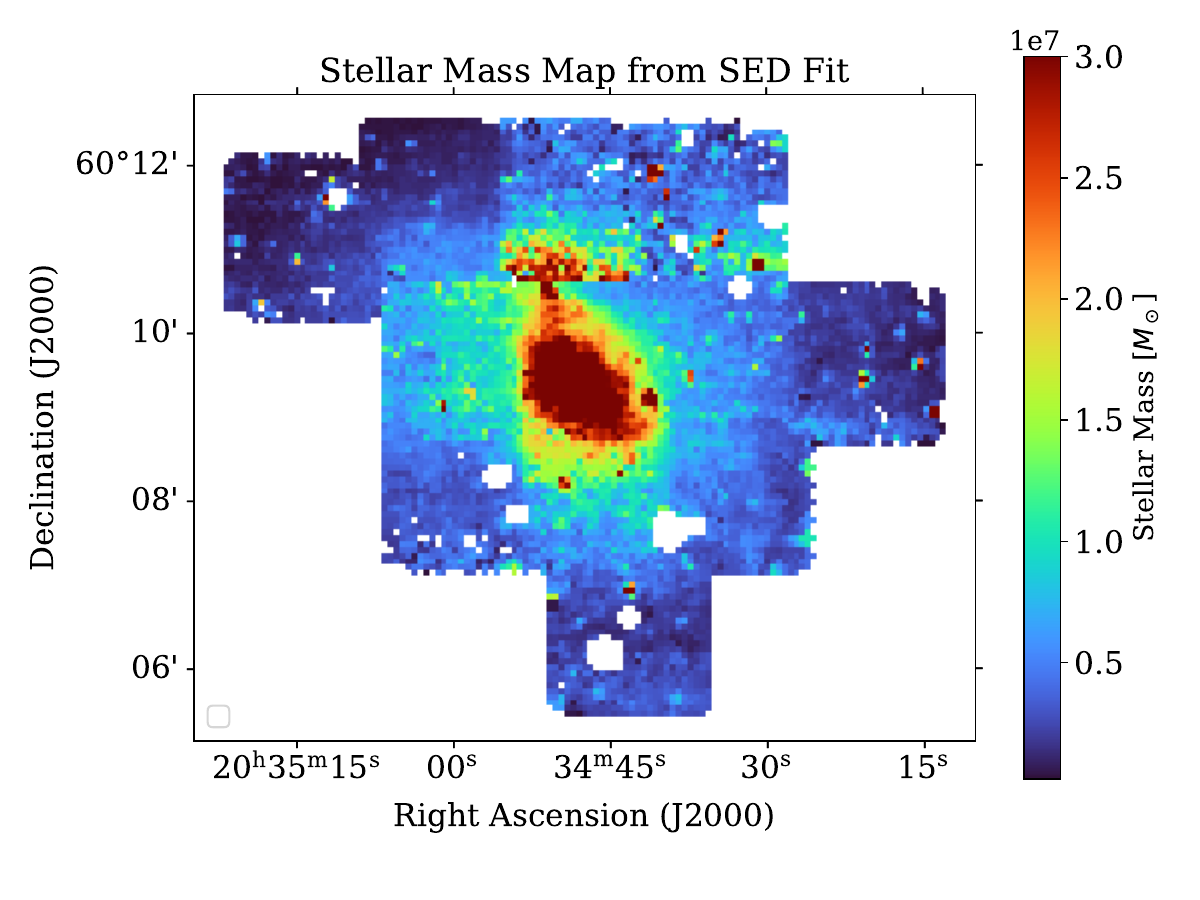}
    \caption{Spatially resolved stellar mass distribution of NGC 6946 from \texttt{SINOPSIS} SED fitting.}
    \label{fig:stellar_mass}
\end{figure}

From the optical spectral fitting, we obtain spatially resolved maps of stellar mass and the luminosity-weighted (LW) and mass-weighted (MW) age for NGC 6946. The stellar mass distribution is presented in Fig.~\ref{fig:stellar_mass}, while the age maps are shown in Fig.~\ref{fig:stellar_ages}. Additional stellar population properties are presented in Appendix~\ref{AppendixA_SSP}.

The total stellar mass we find is  $\log(M_*/M_\odot)=10.74\pm 0.01$. The stellar mass distribution shows strong central concentration, with the bulk of mass contained within approximately 2.6 kpc from the centre, where $\sim 2.11\times10^{10}M_{\odot}$, i.e. almost $40\%$ of the total stellar mass, is located.  \cite{Jarrett} determine a stellar mass for NGC 6946 of $\log(M_*/M_\odot)=10.50\pm 0.09$ (after correcting for differences in the distance), using  data from the Wide-field Infrared Survey Explorer WISE \citep{Wright2010}, specifically the W1 channel at 3.4 $\mu$m, a value of $M_*$ lower than the one we find. An instructive comparison can be made using the total effective radius derived by Valerdi et al. (in prep.) for NGC 6946 in the $K_s$ band, $R_{\mathrm{eff},K_s}=119.76\pm18.50$ arcsec. This value is relatively close to the half-mass radius derived in this work ($\sim92.2$ arcsec; Fig.~\ref{fig:stellar_mass}), indicating that the two estimates trace comparable characteristic scales of the galaxy.

The mean age of stellar populations can be characterised using two distinct weighting schemes: the LW age and MW age. The LW age traces the stellar population that dominate the light and is therefore biased towards young stars, while the MW age reflects the overall stellar content, including older populations. The central region shows $\log(\mathrm{LW\mbox{-}age/yr}) \sim 9$, indicating relatively young stars compared to typical bulges.
The innermost $\sim0.3$ kpc (the region that hosts a nuclear stellar bar, see Section \ref{sec:discussion_atomictomolecular}), present  younger ages, as a consequence of the nuclear SFR burst in this region \citep{Schinnerer2007}. The spiral structure is particularly evident in the LW age map, where green-yellow colours along the spiral arms correspond to $\log(\mathrm{LW\mbox{-}age/yr}) \sim 8.5$--$9$. The youngest populations, with $\log(\mathrm{LW\mbox{-}age/yr}) \leq 8$, are located between $\mathrm{RA} \sim 20^{\mathrm{h}}35^{\mathrm{m}}00^{\mathrm{s}}$--$20^{\mathrm{h}}35^{\mathrm{m}}15^{\mathrm{s}}$ and $\mathrm{DEC} \sim 60^\circ12'$--$60^\circ10'$, co-spatial with regions of recent star formation activity. Some other star-forming regions throughout NGC 6946 share similar characteristics to these.

\subsubsection{Star formation history}
\label{ss: SFH}

\begin{figure}
    \centering
    \includegraphics[width=1\linewidth]{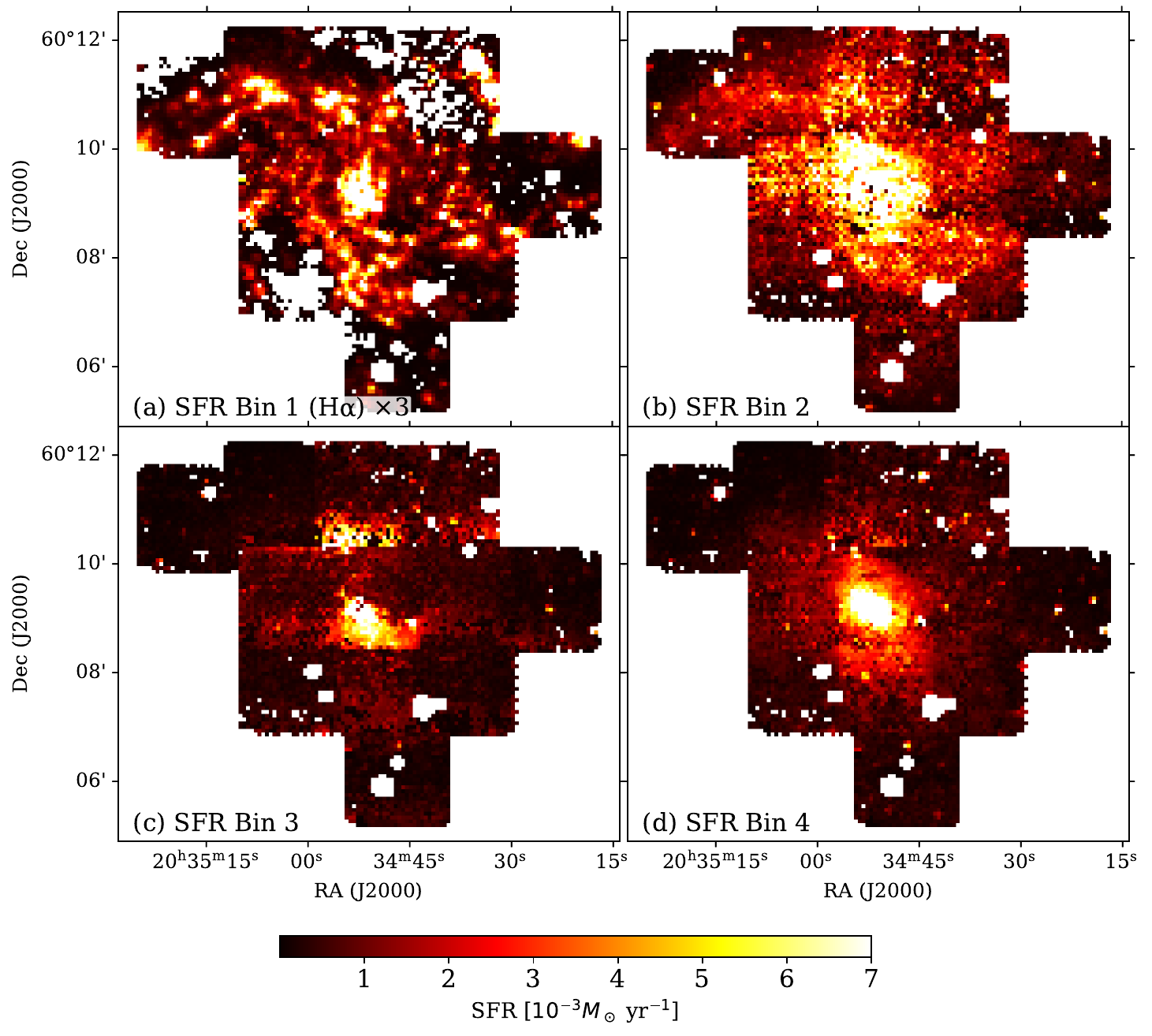}
    \caption{SFR maps for NGC~6946. Panel (a) (Bin~1, H$\alpha$) has been multiplied by a factor of 3 to be visible on the same colour scale. Panels (b), (c), and (d) show the SFR derived from SED fitting using SINOPSIS for older bins.}
    \label{fig:sfr_maps}
\end{figure}

\begin{table*}
\centering
\caption{Age bins used in \texttt{SINOPSIS} and the derived star formation rates.}
\begin{tabular}{lcccc}
\toprule
\textbf{Bin} & \textbf{Age Interval} (yr)& \textbf{SFR} $(M_\odot \, \text{yr}^{-1})$ & \textbf{Method} & \textbf{Notes} \\
\midrule
1 & $0 - 2 \times 10^{7}$ &$3.647 \pm 0.276$ & H$\alpha$ Luminosity & Extinction-corrected
\\
1 & - & $1.389 \pm 0.103$ & H$\alpha$ Luminosity & Non-Extinction-corrected

\\
1-2$^a$& - & $3.235 \pm 0.100$ & Infrared Luminosity & $8 - 1000 \, \mu\text{m}$
\\
2 & $2 \times 10^{7} - 5.7 \times 10^{8}$ &$14.63$ [$3.93$,  $22.07$] & \texttt{SINOPSIS} SED Fitting & Range from 11 realisations \\
3 &  $5.7 \times 10^{8} - 5.7 \times 10^{9}$&$5.54$ [$1.34$,  $9.60$] & \texttt{SINOPSIS} SED Fitting & Range from 11 realisations \\
4 & $5.7 \times 10^{9} - 1.4 \times 10^{10}$ &$7.03$ [ $3.15$, $11.36$] & \texttt{SINOPSIS} SED Fitting & Range from 11 realisations \\
\bottomrule
\end{tabular}
  \tablefoot{$^{(a)}$~This SFR tracer accounts for $\sim100$ Myr, corresponding to the initial part of age bin 2, which extends to $\sim570$ Myr.
    }
\label{tab:sfr_values}
\end{table*}
The recent SFR ($\sim 10$ Myr) was estimated through the H$\alpha$ emission (Section \ref{sec:ionized_gas}), but the FIR (8-1000$\mu$m) can also be used as an alternative calibrator \citep{Kennicutt1998}, which traces star formation over a longer timescale ($\sim 100$ Myr). We used the calibration from \cite{Hayward2014} for isolated disc galaxies, while $L_{\text{IR}}$ was obtained by integrating the SED of the MBB model with the previously fitted parameters. \texttt{SINOPSIS}  allows us to obtain the SFR in three additional bins of stellar age. \texttt{SINOPSIS} also returns uncertainties on the derived quantities by providing their maximum and minimum values across multiple realisations of the fit; these values are then used to compute error bars. The values for bins 2, 3, and 4 were derived from the \texttt{SINOPSIS} SED fitting, where the range reflects the minimum and maximum values from the 11 realisations, enough to  explore well the best models parameters space (see \citealt{Fritz2007} for details). Fig.~\ref{fig:sfr_maps} presents the SFR maps across the four distinct age bins, with corresponding values provided in Table~\ref{tab:sfr_values}. These maps reveal the underlying cause for the central concentration of stellar mass observed in Section~\ref{ss:stellar mass and age}. Prominent star formation bursts are evident in all bins, with the star formation activity in bin 2 being distributed throughout the whole disc, while in the third bin it is strongly concentrated within $\sim2$ kpc (consistent with the stellar mass distribution in Fig.~\ref{fig:stellar_mass}): in fact, this pattern is fully consistent with the mean stellar age distributions  previously presented. A significant star formation downturn occurs during bin 1, preceded by a remarkable burst in bin 2 that reaches the highest SFR values in the SFH of NGC 6946 (see Table~\ref{tab:sfr_values}). The galaxy-wide nature of this bin 2 burst suggests a global triggering mechanism, which we will investigate by examining the ISM properties in subsequent sections, focusing on dust, HI, and H$_2$ characteristics to identify potential causes and consequences.

\subsection{ISM properties}
\subsubsection{Ionised gas}
\label{sec:ionized_gas}

The gas kinematic derived from the emission lines fitting reveals a clear rotation pattern in which one part is receding (south-west) and the other is approaching  the observer. This is visible on both the stellar and the ionised gas velocity maps, indicating that the two  components share the same velocity pattern, the classical rotation pattern of spiral galaxies. Fig. \ref{fig:sfr_maps}a presents the H$\alpha$ emission. The pattern of the spiral arms is clearly distinguishable, reflecting the fact that recent star formation in the disc is primarily concentrated in these regions.\\
We also produced the BPT diagram \citep{Baldwin1981}, showing that SF is the dominant ionisation mechanism across NGC 6946 and with no signs of an AGN. Our diagram agrees with the presented by \cite{MetalTHINGS}. More notably, we have detected the ultraluminous X-ray source ULX-3, which is  visible in the joint \textit{XMM-Newton} and NuSTAR observations, located at the coordinates (J2000) 20:35:00.7 +60:11:31 \citep{Liu2005, Earnshaw2019, Fabbiano1987}. This source, located on the northern side of the galaxy, clearly exhibits accreting black hole powered emission lines, falling within the AGN and composite regions of the BPT diagrams. This classification is reliable and not affected by low S/N or by artefacts.  We plan to complement the study of NGC 6946 -ULX-3 with our dataset in a future work.\\
The conversion from $L_{H\alpha}$ to SFR was performed using the calibrated factor from \cite{Kennicutt1998} for a \cite{Chabrier2003} IMF using only the spaxels classified as star forming  and  composite in the BPT diagram, which amount to approximately $\sim 97\%$ of the total. The recent SFR using this method, both with and without accounting for dust extinction correction, is $3.65\pm 0.28\, M_\odot \, \text{yr}^{-1}$ and $1.39\pm 0.10\, M_\odot \, \text{yr}^{-1}$, respectively. At this point, it is important to mention a caveat. The central kiloparsec of NGC 6946 is known to host large amounts of dust and strong extinction  \citep[][but see also Section \ref{sec:dust}]{Schinnerer2006}, making accurate SFR measurements from optical recombination lines challenging. Although we corrected the H$\alpha$ emission using the Balmer decrement, this approach is, by nature, not sensitive to ionising photons originating from deeply embedded star-forming regions where optical emission is heavily attenuated. \cite{Eibensteiner2022} measure a mean $\Sigma_{\rm SFR}\sim0.79\ \mathrm{M_\odot\,yr^{-1}\,kpc^{-2}}$ for the inner 1 kpc using the 33 GHz free-free continuum, an extinction-free tracer of recent massive star formation, whereas our extinction-corrected H$\alpha$ map yields $\Sigma_{\rm SFR}\sim0.32\ \mathrm{M_\odot\,yr^{-1}\,kpc^{-2}}$ over the same region. The comparison suggests that the optical tracer likely underestimates the true SFR in the nucleus due to extremely strong obscuration. Consequently, the central SFR values presented here should be regarded as conservative estimates, or lower limits. This limitation is expected to primarily affect the innermost region and does not alter the main conclusions regarding the star formation properties across the galaxy disc. As a consequence, we expect the most central points in the resolved KS relation to shift towards higher $\Sigma_{\rm SFR}$ when using extinction-free tracers.

We also computed the oxygen abundance for SF spaxels   in the BPT-NII diagram. Given the available spectral range, we employed the three-dimensional (3D) calibration O/H $= f(N2, S2, R3/S2)$ from \citet{Pilyugin2016}, which comprises two regimes corresponding to high and low metallicity. A Monte Carlo method was used to propagate the flux uncertainties into the final metallicity values. We fit a linear relation to the radial metallicity gradient, yielding
\begin{equation}
    12+\log(\text{O/H}) = (8.614 \pm 0.016) - (0.1304 \pm 0.0122)\,\frac{R}{R_{\text{eff}}},
    \label{eq:OH_grad}
\end{equation}
where $R_\text{eff}$ is the near-ultraviolet (NUV) effective radius (Valerdi et al. in prep). \cite{Vale2025} use the same data and the same prescription to calculate spatially resolved maps for the gas metallicity, but with different analysis tools. The slope and intercept values we find match their results ($-0.104\pm0.002$ and $8.599\pm0.002$, respectively),  displaying a relative difference of less than $\sim 25 \%$ in the slope, while the intercepts agree within uncertainties. This agreement is quite remarkable, in view of the differences between the way data were analysed. In particular, \citet{Vale2025} employed a different SED fitting code (\texttt{STARLIGHT}; \citealt{STARLIGHT}), a distinct stellar population base \citep{BC2003}, different metallicities ($Z = 0.005, 0.02, 0.05$), a different attenuation law \citep{Cardelli1989}, and a different emission line fitting code \citep{Zinchenko2024}. The fact that such distinct configurations (see Section~\ref{ConfSINOPSIS} for the details of our \texttt{SINOPSIS} setup) yield fully compatible results, reinforces the reliability of the result, suggesting that -- at least under the assumption for the metallicity calculation -- this is a robust and stable result, independent of the specific adopted fitting tools.

\subsubsection{Dust}
\label{sec:dust}

In this section, we analyse dust in its two observable forms: absorption and emission. Fig. \ref{fig:ext_sinopsis} presents the extinction map derived from the SED fitting with \texttt{SINOPSIS}, relative to stellar populations dominating the continuum emission (i.e. age $>2\times 10^7$ yr). The extinction obtained via \texttt{SINOPSIS} corresponds to the effective stellar attenuation (i.e. how dust reduces the stellar flux when fitting the full spectrum). It is modelled under the assumption of a uniform dust screen in front of the stars and  parametrised as $A_V$ using the extinction law from \cite{Calzetti1994}. Although the physical distribution is more complex, with dust and stars coexisting in the same regions, a more realistic description would require treatment with radiative transfer codes. On the other hand, Fig. \ref{fig:contour_overview}b shows the extinction derived from the Balmer decrement from the $H\alpha$ and $H\beta$ emission lines. An increase in extinction towards the galactic centre is observed in extinction maps. This is the effect of the higher concentration of dust in the central region, possibly associated with both:  past episodes of star formation (see Sections \ref{ss: SFH}  and \ref{sec:dust}), which could have enriched the ISM with dust through SNe and evolved stars, and the ISM inflow towards the centre due to the stellar bars that NGC 6946 hosts. 

Regarding radiation, dust emits almost all of its continuum radiation in the FIR. At long wavelengths ($\lambda>50\mu$m), the emission is dominated by dust grains in thermal equilibrium with the surrounding radiation field, at temperatures usually in the $\sim 15-35$ K range, contributing about $\sim2/3$ of the total infrared (IR) emission, while the remaining $\sim 1/3$ originates from polycyclic aromatic hydrocarbons (PAHs) \citep{Draine2003} and small dust grains, which are not in thermal equilibrium with the local radiation field and dominate emission at shorter wavelengths. The total background-corrected fluxes for each {\it Herschel} band are presented in Table \ref{tab:fluxes}. We find that the uncertainties are dominated by calibration errors in all bands, with instrumental and background uncertainties contributing negligibly.

We used a MBB fitting (Equation~\ref{eq:mbb}) to model the SED in the 100 to 500 $\mu$m range, letting the free parameters be $M_d$, $M_T$, and $\beta$. Uncertainties on these values were calculated by bootstrap on 5000 different realisations of the observed SED. The 70~$\mu$m band was excluded from the fit as it systematically lies above the model prediction. This discrepancy suggests the presence of a warmer dust component not captured by the single-temperature MBB model. The fitted dust temperature of $\sim$22~K corresponds to thermal emission from cold dust, which is insufficient to account for the observed 70~$\mu$m flux, a result consistent with the findings of similar cases  \citep[see e.g.][]{Smith2010}. While an additional MBB component would be required to model this emission, this would be poorly constrained due to having only a single data point at this wavelength. However, the bulk of the dust mass is dominated by the cold component \citep[e.g.][]{Kramer2010}.
The five-band fit results are summarised in Table \ref{tab:fit_results}.

\begin{table}
\centering
\caption{Integrated results from the MBB fit to the \textit{Herschel} data.}
\begin{tabular}{lc}
\hline
\textbf{Parameter} & \textbf{Value} \\
\hline
Dust Mass, $M_d$ & $(9.525 \pm 0.870) \times 10^7 \, M_\odot$ \\
Dust Temperature, $T_d$ & $21.85 \pm 0.86$ K \\
Emissivity Index, $\beta$ & $2.072 \pm 0.126$ \\
Median $\chi^2$ & $1.86$ \\
\hline
\end{tabular}
\label{tab:fit_results}
\end{table}

Despite the relatively poor spatial resolution of the FIR data, especially for the SPIRE instrument, the proximity to this galaxy allows us to model the fluxes on a pixel-by-pixel basis. To accomplish this task, we imposed quality criteria requiring that each pixel must have valid flux measurements in at least three wavelengths: the fluxes need to be exceeding $2\sigma_{\text{noise}}$, where $\sigma_{\text{noise}}$ was calculated as the standard deviation of pixels in sky regions. Pixels meeting these criteria were included in the fitting process. The resulting spatially resolved maps of $M_d$, $T_d$, and $\beta$ are presented in Fig.~\ref{fig:dust_properties_5pt}.

Summing the dust mass across all valid pixels and considering a conservative uncertainty of $\sim$14\% based on the error propagation from the global fit, we obtain a total dust mass value of $M_d = (1.027 \pm 0.144) \times 10^{8}~M_{\odot}$. This value is approximately 9\% higher than the result of the global integrated flux analysis.

To characterise the dust emissivity index for NGC 6946, we computed the distribution of $\beta$ values across the galaxy. The median value of this distribution, $\beta \sim 2.1$ was adopted as a representative global dust emissivity index for the subsequent analysis. With the aim of improving the mapping of dust emission, we now turn to the simplified approach already outlined in Section \ref{methods:dust}, where only two bands are exploited.
While this method sacrifices spatially resolved information about the emissivity index $\beta$, we mitigated this limitation by establishing a characteristic value of $\beta = 2.1$ from the  analysis of the $\beta$ distribution across NGC 6946. With the former in mind, adopting $\beta = 2.1$ as characteristic of the entire galaxy -- consistent also with the value derived from the five-point global fit ($\beta = 2.072$) -- we applied Equation~\ref{eq:flux_ratio} to numerically solve for temperature using the 70 and 100~$\mu$m total fluxes. This two-band approach provides higher spatial resolution for resolved analysis ($6.7\arcsec \times 6.9\arcsec$ i.e. $\sim 258$ pc). With the temperature determined, we directly computed the dust mass using Equation~\ref{eq:mbb} with either flux measurement. This method yields $M_d = (3.65 \pm 0.76) \times 10^{7}~M_{\odot}$ and $T_d = 25.55 \pm 0.76$~K for $\beta = 2.1$. The derived dust mass is of the same order of magnitude as previous estimates from global and spatially resolved five-point fits, though  lower. This discrepancy can be explained by the use of 70~$\mu$m data, which traces warmer dust components that, because of the higher temperature, emit more efficiently per unit mass. 

\begin{figure}
    \centering
    \includegraphics[width=0.99\linewidth]{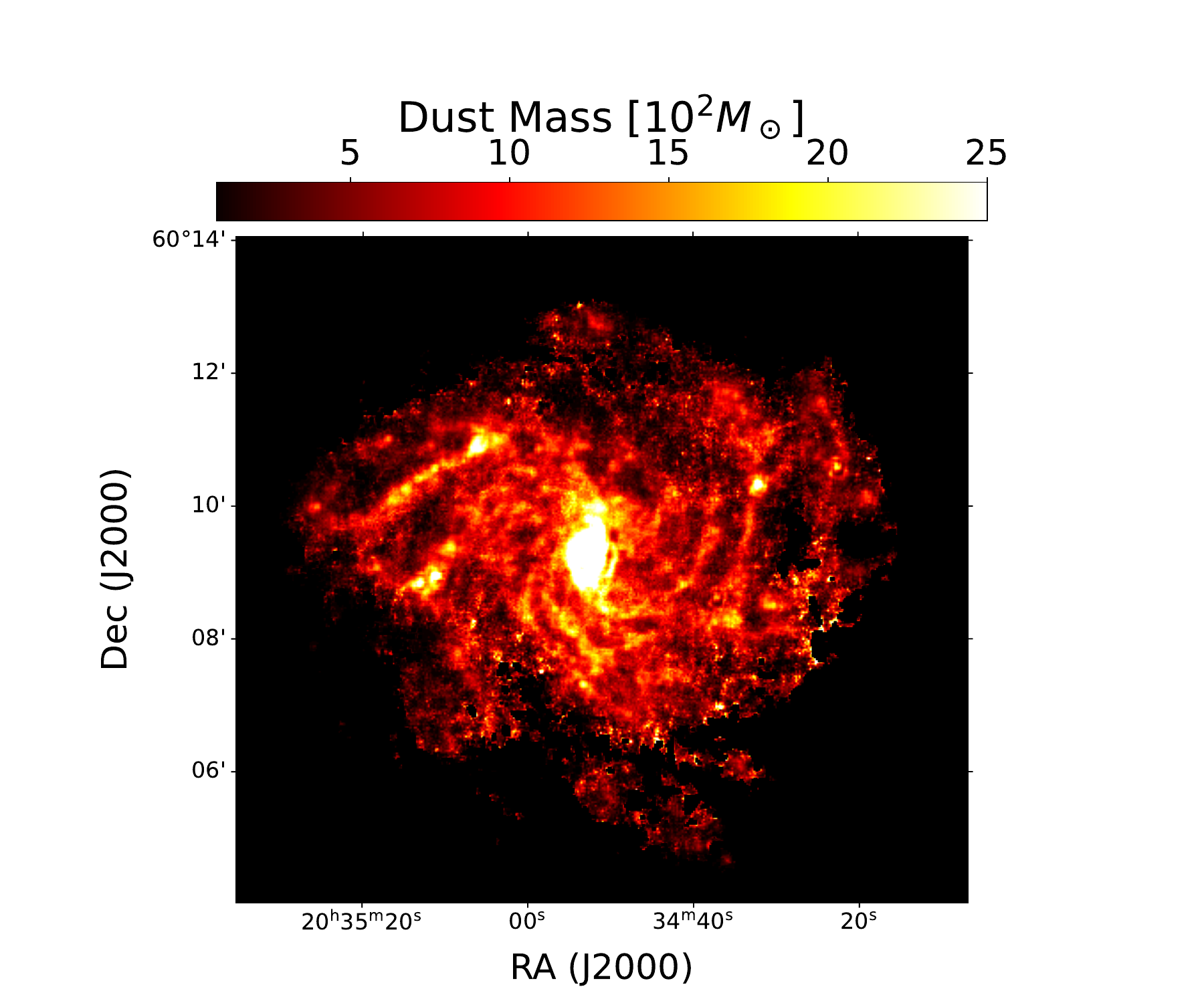}
    \caption{Spatially resolved dust mass of NGC 6946 derived from 70 and 100~$\mu$m \textit{Herschel} images.}
    \label{fig:dust2pt}
\end{figure}

Applying the same two-point  procedure to each pixel using the 70~$\mu$m image convolved and reprojected to match the 100~$\mu$m map resolution, we generated a new spatially resolved map for dust mass at significantly higher resolution (Fig.\ref{fig:dust2pt}). The dust mass obtained from different methods are presented in Table \ref{tab:dust_properties} (appendix \ref{AppendixB_dust}).  The spatially resolved two-point fit map  will be used only to create dust mass contours (Section \ref{ss:gas_dust_trends}), and hence only address the dust morphology. This approach is chosen because the 70-100 $\mu$m images allow us to recover dust geometry at a higher resolution.

The derived dust masses show systematic variations across methods. The spatially resolved five and two-point fits yields higher masses than their global counterparts. We attribute this to the fact that the pixel-by-pixel analysis  better captures faint emission that contributes to the total mass but is under-represented in integrated flux measurements. The two-point fit yields lower masses but higher temperatures, reflecting the bias towards warmer dust components when including 70 $\mu$m emission. Warmer dust emits more efficiently, requiring less mass to produce the observed fluxes.
All of our $M_d$ estimates are consistent with  values found in the literature, provided that the systematic differences between dust models are considered. For instance, in DustPedia \citep{Davies2017}, NGC 6946 is reported to have a total dust mass of $(2.61 \pm 0.47)\times10^7 M\odot$. This value is lower than our estimates, but this discrepancy is expected given that DustPedia uses the CIGALE code \citep{Boquien2019} based on the THEMIS dust model \citep{Jones2017}. Such differences are well documented: the THEMIS model typically yields dust masses that are a factor of 2 to 3 lower than those derived from other models like \cite{Draine_Li2007}, as discussed in \cite{Casasola2020} and \cite{Pantoni2026}.

We also compare our dust mass estimate with the results of \citet{Aniano2012}, who used the \citet{Draine_Li2007} dust models to interpret  the observed broad band SED from 3.6 to 500~$\mu$m together. Our results are in very good agreement, notwithstanding their finding that dust masses derived from full SED modelling tend to be systematically larger than those obtained from single-temperature MBB fits. As discussed above, the bulk of the dust mass is primarily constrained by the FIR emission between 100 and 500~$\mu$m. This approach has been also validated by \citet{Bianchi2013}, who showed that reliable dust masses can be obtained either from full dust SED modelling or from simpler single-temperature MBB fits over the 100--500~$\mu$m range.

\subsubsection{Atomic and molecular gas}
We obtain a total atomic gas mass of $8.13\pm0.71\times10^9 \, \text{M}_\odot$, in agreement with \cite{Walter2008} after accounting for distance differences  and a total molecular gas mass of $7.49\pm2.43\times10^9 \, \text{M}_\odot$ (without  helium contribution). This results in a comparable molecular and atomic gas mass content, but with the molecular gas being twice as massive as the atomic gas within $R_{25}$ (see Table~\ref{tab:mass_components}). The total values yield $M_{\mathrm{HI}}/M_{\mathrm{H_2}} \sim 1.1$, consistent with \cite{Leroy2009}. These results indicate that, although HI generally dominates the cold gas content in star-forming galaxies, NGC~6946 represents a case where molecular gas production is enhanced. The respective HI and H$_2$ maps are presented in Appendix \ref{AppendixC_gas}. The optical disc ($R_{25}$) is overlaid on both figures. From now on, we restrict our analysis of global properties to the $R_{25}$ disc, with the mass components summarised in Table~\ref{tab:mass_components}. The choice of adopting $R_{25}$ is  commonly adopted (\citealp[e.g.][]{Casasola2020}): this normalisation ensures that various tracers are compared across the same galactic regions, providing a consistent spatial framework for our results. The HI disc extends significantly beyond the $R_{25}$ radius, exceeding the spatial distribution of the stellar, molecular, and dust components.

According to the DustPedia classification \citep{Makarov2014, Bianchi2018}, NGC 6946 is characterised by a morphological stage $T = 5.9$, placing it within the $T = 6$ sample bin. For this morphological type, \citet{Casasola2020} report an average ratio within $R_{25}$ of $M_{\mathrm{HI}}/M_{\mathrm{H}_2} = 0.96 \pm 0.31$. In contrast, NGC 6946 exhibits a significantly lower value of $\sim 0.46$, indicating a particularly high efficiency in converting atomic gas into molecular form. We associate this behaviour with the combined effects of the bar and the dust content, as discussed in more detail in Section~\ref{sec:discussion_atomictomolecular}.

Our derived gas-to-dust ratio for NGC 6946, GDR$ = 102.0 \pm 22.9$, is consistent with Milky Way measurements (\citealp[typically 100--150; e.g.][]{Elia2017}). Our result is also in good agreement with the average value for the DustPedia $T = 6$ sample, GDR$ = 202 \pm 105$ \citep{Casasola2020}, when adopting a constant $X_{\text{CO}}$ and excluding the contribution of helium to ensure a consistent comparison. While our value is lower than the morphological average, it remains well within the $1\sigma$ scatter. On the other hand, this highlights the remarkable nature of the central region of NGC 6946, where the GDR ratio in the innermost region exceeds the mean value in the rest of the disc by more than a factor of two (see Section~\ref{sec: rings}).

\begin{table}
    \centering
    \caption{Stellar and ISM components within $R_{25}$ for NGC 6946.}
    \label{tab:mass_components}
    \begin{tabular}{lc}
        \hline
        Component & Mass ($M_{\odot}$) \\
        \hline
        Stellar Mass & $(5.59 \pm 0.06) \times 10^{10}$ \\
        Dust Mass & $(1.03 \pm 0.05) \times 10^{8}$ \\
        HI Mass & $(3.30 \pm 0.29) \times 10^{9}$ \\
        H$_2$ Mass & $(7.17 \pm 2.33) \times 10^{9}$ \\
               Total gas HI $+$H$_2$ Mass & $(1.05 \pm 0.23) \times 10^{10}$ \\
               Gas-to-Dust Ratio & $(102.00 \pm 22.87) $ \\
        \hline
    \end{tabular}
\end{table}

\subsection{Star-forming main sequence}

As a comparison with our values of $M_*$ and recent SFR, we used the relation derived by \cite{Cano2016A}, who construct the star-forming main sequence (SFMS) \citep{Brinchmann2004} from spatially resolved observations of the CALIFA survey \citep{CALIFA2012}. They integrate the  properties of 306 local galaxies to obtain their global values. The sample includes only those galaxies whose ionisation is dominated by young stars, based on the BPT criteria and the H$\alpha$ EW. The galaxies in the sample span a redshift range of $0.005<z<0.03$.
The position of this galaxy on the SFMS is shown in Fig.~\ref{fig:sfms}. The figure plots recent SFRs derived using two distinct tracers and timescales: (1) the extinction-corrected H$\alpha$ luminosity (blue square), which traces star formation over a timescale of $\sim 10$ Myr, and (2) the total SFR from the combination of unobscured (non-extinction-corrected H$\alpha$) and obscured (infrared luminosity) emission (red square), corresponding to a timescale of $\sim 100$ Myr (see Table~\ref{tab:sfr_values} for individual values). These results about the position of NGC 6946 in the SFMS are particularly important for our analysis (Sections \ref{sec:discussion}, \ref{sec:conclusions}). It is important to mention that CALIFA relies only on $H\alpha$ for the SFR determination and that the FIR traces a slightly longer SFR timescale ($10^8$ yr).

\section{Radial analysis}
\label{sec: rings}
\begin{figure}
    \centering
    \includegraphics[width=0.9\linewidth]{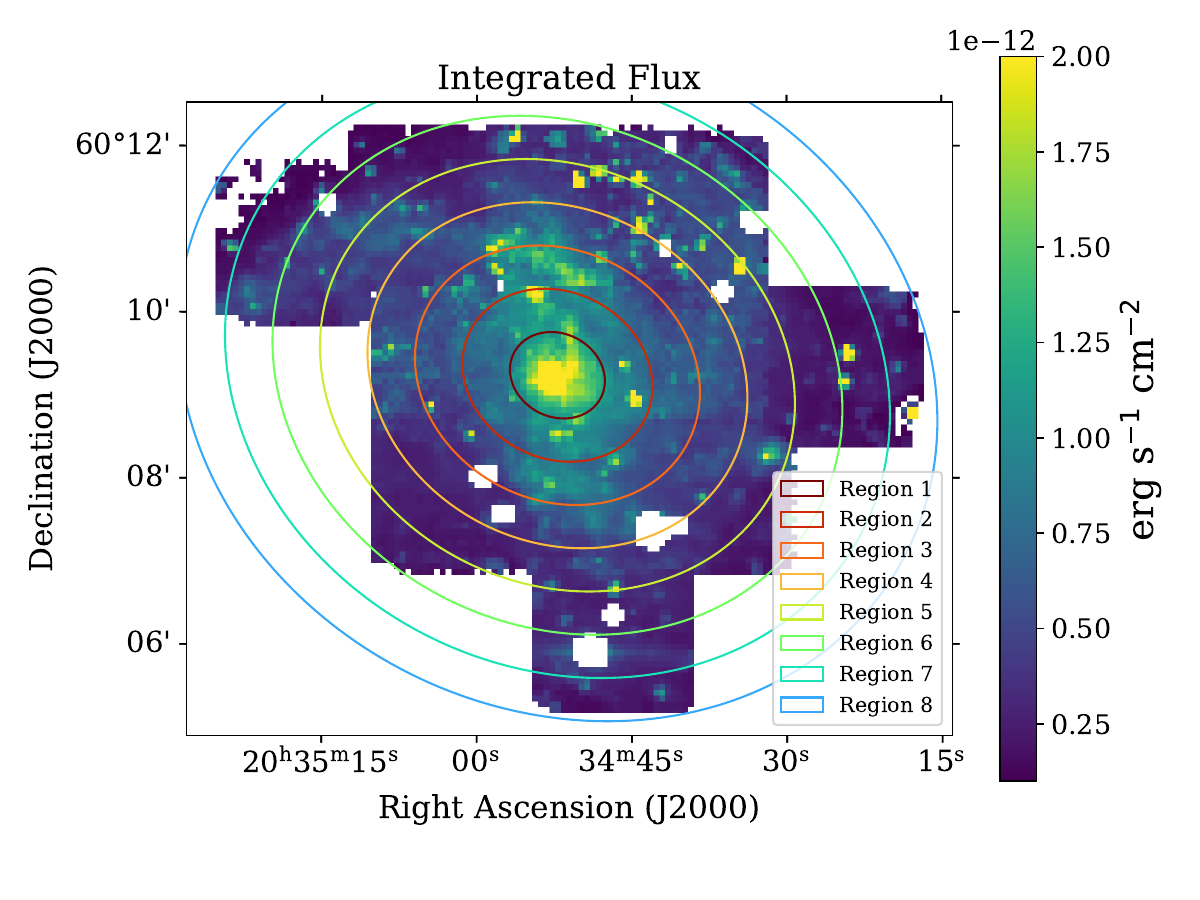}
    \caption{Eight concentric elliptical regions used for radial analysis, sharing identical orientation and axis ratio. Each pixel is assigned to a single region.}
    \label{fig:regions}
\end{figure}

For a quantitative radial analysis, we divided NGC 6946 into eight concentric elliptical regions sharing identical axis ratios and orientation, to perform a tilted ring analysis. The ellipse P.A. (= $242.7^\circ$) follows \cite{deBlok}, measured from north to east along the receding side. Sizes were normalised to $R_{25}$ and axis ratio from \cite{Karachentsev2004}: $a = 11.5\,\text{arcmin}, b/a = 0.85$. The outermost ellipse encompasses the whole extent of our optical data, corresponding to a physical semi-major axis $a \sim 10.7$\,kpc for the physical scale from the adopted distance (155.6\,pc\,spaxel$^{-1}$). The eight concentric ellipses are shown in Fig.~\ref{fig:regions} and Table \ref{tab:ellipse_axes}.

\begin{table}
    \centering
    \caption{Semi-major axis ($a$) in arcseconds and kiloparsecs for the eight elliptical regions used in the tilted-ring analysis.}
    \label{tab:ellipse_axes}
    \begin{tabular}{ccc}
        \hline
        Region  & $a$ (arcsec) & $a$ (kpc) \\
        \hline
        1  & 35.0 & 1.33 \\
        2  & 70.0 & 2.65 \\
        3  & 105.3 & 4.00 \\
        4  & 140.6 & 5.34 \\
        5  & 175.9 & 6.68 \\
        6  & 211.1 & 8.02 \\
        7  & 246.4 & 9.35 \\
        8  & 281.7 & 10.69 \\
        \hline
    \end{tabular}
\end{table}

\subsection{Star formation history trends}
\begin{figure}
    \centering
    \includegraphics[width=0.8\linewidth]{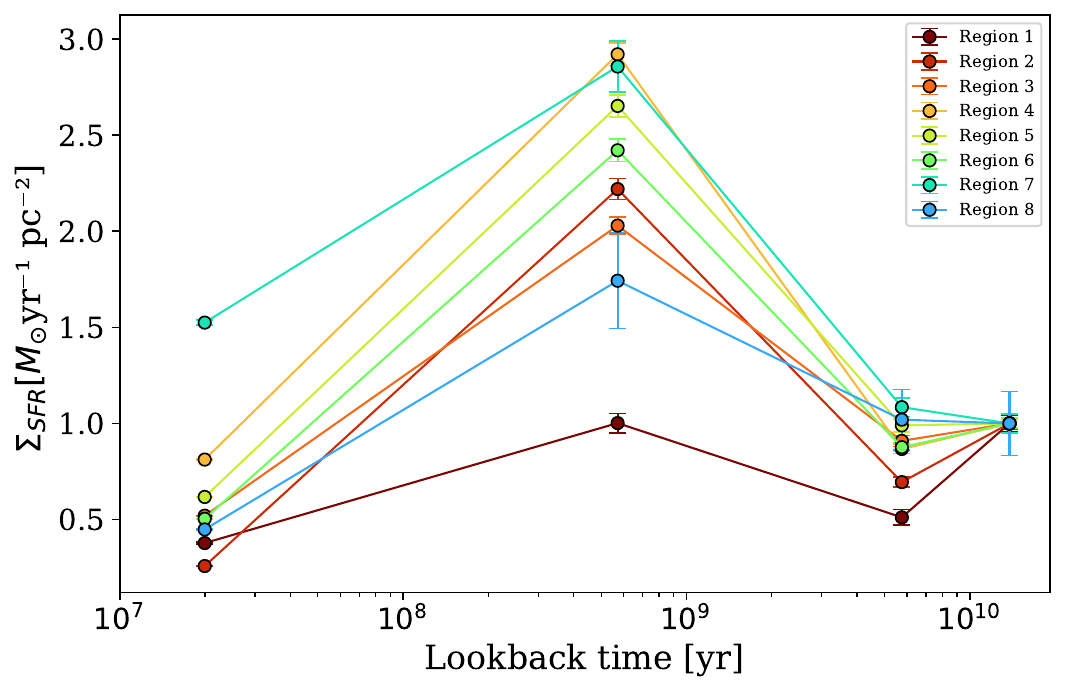}
    \caption{$\Sigma_\text{SFR}$ for each elliptical region, normalised to the corresponding bin 4 values. This normalisation highlights the temporal evolution of star formation intensity relative to the earliest epoch.}
    \label{fig:SSFR_normalized}
\end{figure}

\begin{figure}
    \centering
    \includegraphics[width=0.95\linewidth]{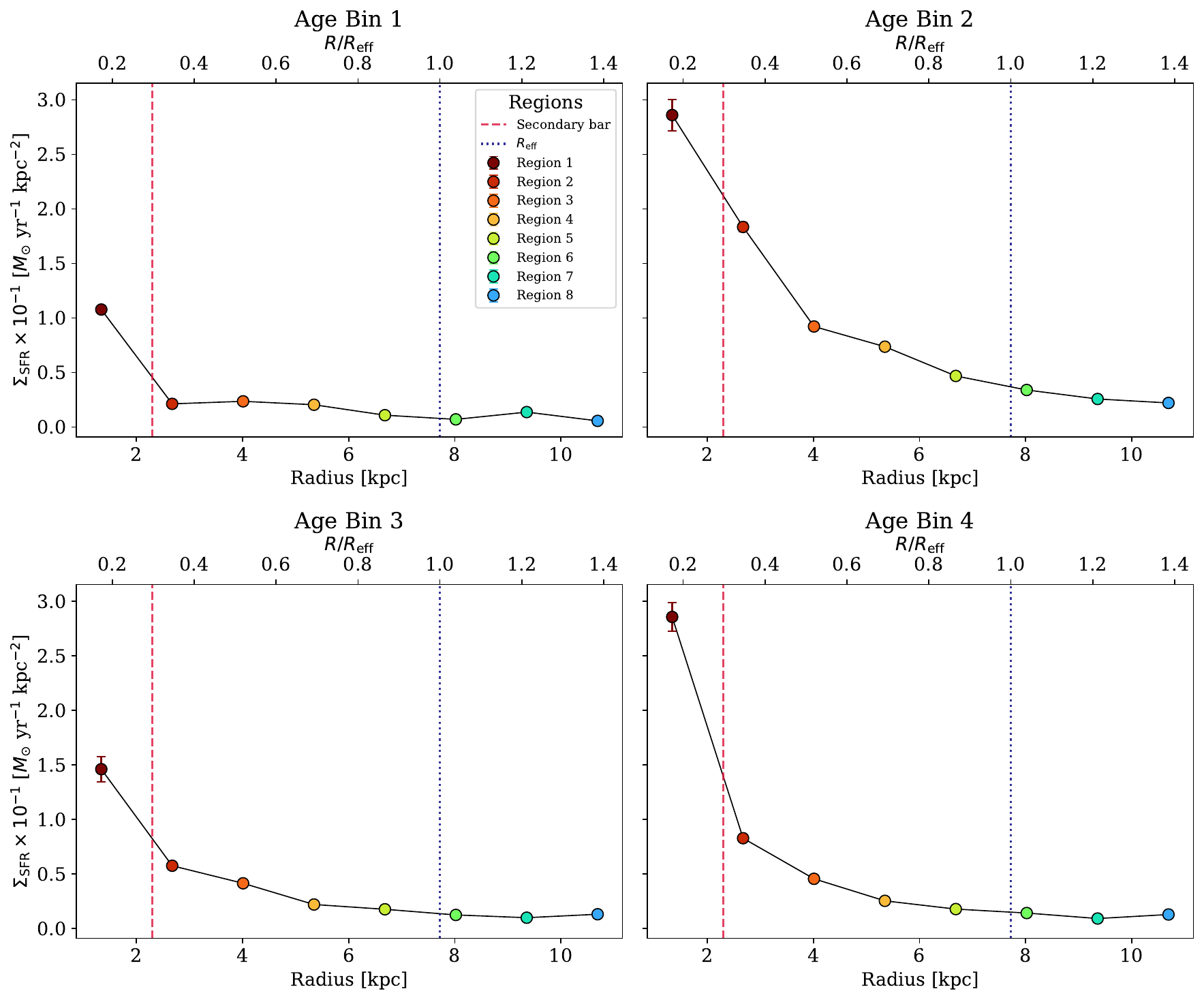}
    \caption{$\Sigma_\text{SFR}$ as a function of semi-major axis for each age bin. The vertical dashed red line corresponds to the radius of the secondary bar, while the blue line corresponds to $R_{\text{eff}}$.}
    \label{fig:SSFR_R}
\end{figure}

Radial SFR measurements are presented in Table~\ref{tab:sfr_regions}. Fig.~\ref{fig:SSFR_normalized} displays the evolution of the star formation rate surface density ($\Sigma_\text{SFR}$) normalised to its value in age bin 4, revealing the relative changes in star formation intensity across different galactic regions over time and where is easy to see when the SFR peak occurred in all regions. Another view of this plot is given in Fig.~\ref{fig:SSFR_R}, where the actual $\Sigma_\text{SFR}$ values are presented as radial trends for each age bin. The $\Sigma_\text{SFR}$ radial profiles reveal distinct evolutionary  patterns as a function of the galactocentric distance: all age bins exhibit negative radial gradients except bin 1, which shows mild enhancements in regions 3 and 7, and is overall the one presenting the flattest trend. Region 1 (within 1.33 kpc) consistently maintains the highest $\Sigma_\text{SFR}$ values across all epochs, with the absolute maximum occurring in age bin 2, where the highest total SFR for NGC 6946 is found. 

\begin{table}
    \centering
    \caption{$\Sigma_{\mathrm{SFR}}$ in $\times 10^{-3} M_{\odot}\,\mathrm{yr}^{-1}\,\mathrm{kpc}^{-2}$ by elliptical region for four age bins.}
    \label{tab:sfr_regions}
    \begin{tabular}{ccccc}
        \hline
        R & Bin 1 (H$\alpha$) & Bin 2 & Bin 3 & Bin 4 \\
        \hline
        1 & $108 \pm 1.15$ & $286 \pm 14.3$ & $146 \pm 11.5$ & $286 \pm 13.0$ \\
        2 & $21.3 \pm 0.09$ & $184 \pm 4.56$ & $57.4 \pm 2.15$ & $82.7 \pm 2.27$ \\
        3 & $23.6 \pm 0.08$ & $92.3 \pm 2.04$ & $41.3 \pm 1.29$ & $45.4 \pm 0.99$ \\
        4 & $20.5 \pm 0.07$ & $73.7 \pm 1.47$ & $21.9 \pm 0.58$ & $25.2 \pm 0.54$ \\
        5 & $10.9 \pm 0.05$ & $46.9 \pm 1.01$ & $17.5 \pm 0.60$ & $17.7 \pm 0.46$ \\
        6 & $7.08 \pm 0.03$ & $34.0 \pm 0.81$ & $12.3 \pm 0.31$ & $14.1 \pm 0.38$ \\
        7 & $13.8 \pm 0.12$ & $25.8 \pm 1.19$ & $9.78 \pm 0.44$ & $9.02 \pm 0.46$ \\
        8 & $5.70 \pm 0.05$ & $22.1 \pm 3.16$ & $13.0 \pm 2.00$ & $12.7 \pm 2.11$ \\
        \hline
    \end{tabular}
\end{table}

\begin{figure*}
  \centering
  {
    \includegraphics[width=0.32\textwidth]{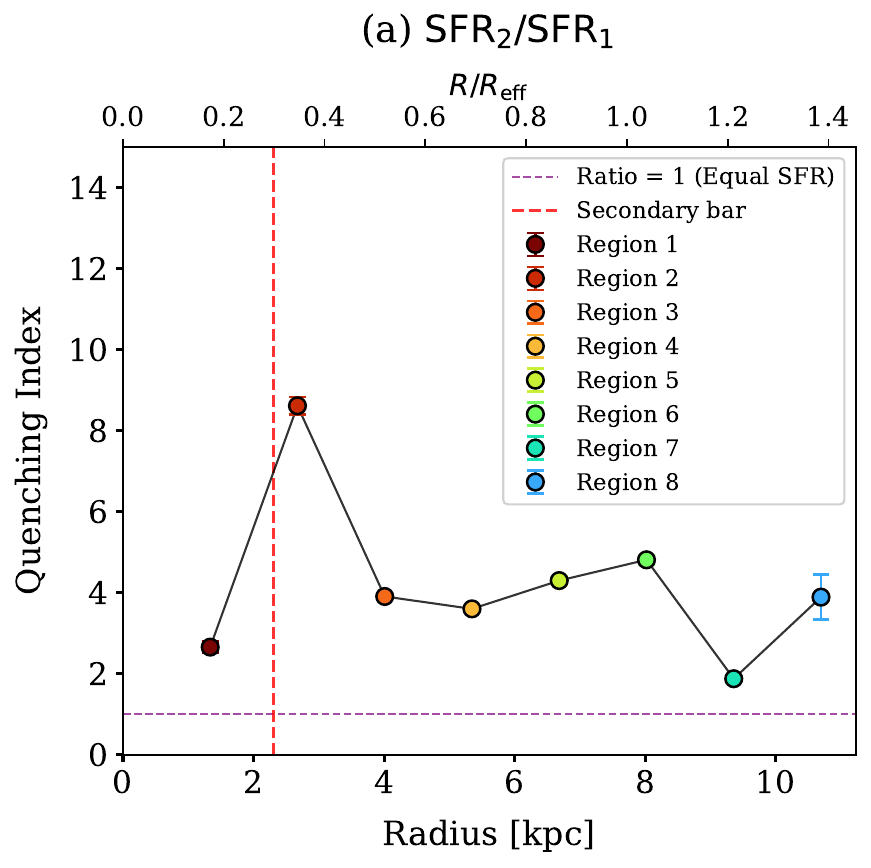}
    \label{fig:SFR_ratio_1}
  }
  \hfill
 {
    \includegraphics[width=0.32\textwidth]{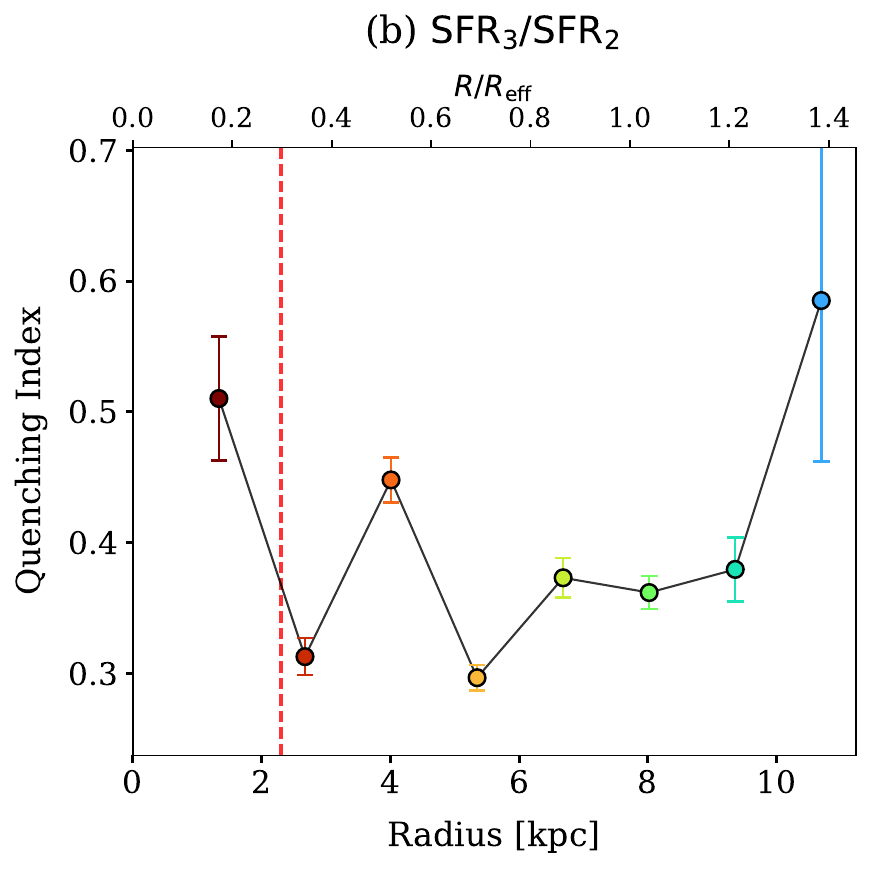}
    \label{fig:SFR_ratio_2}
  }
  \hfill
  {
    \includegraphics[width=0.32\textwidth]{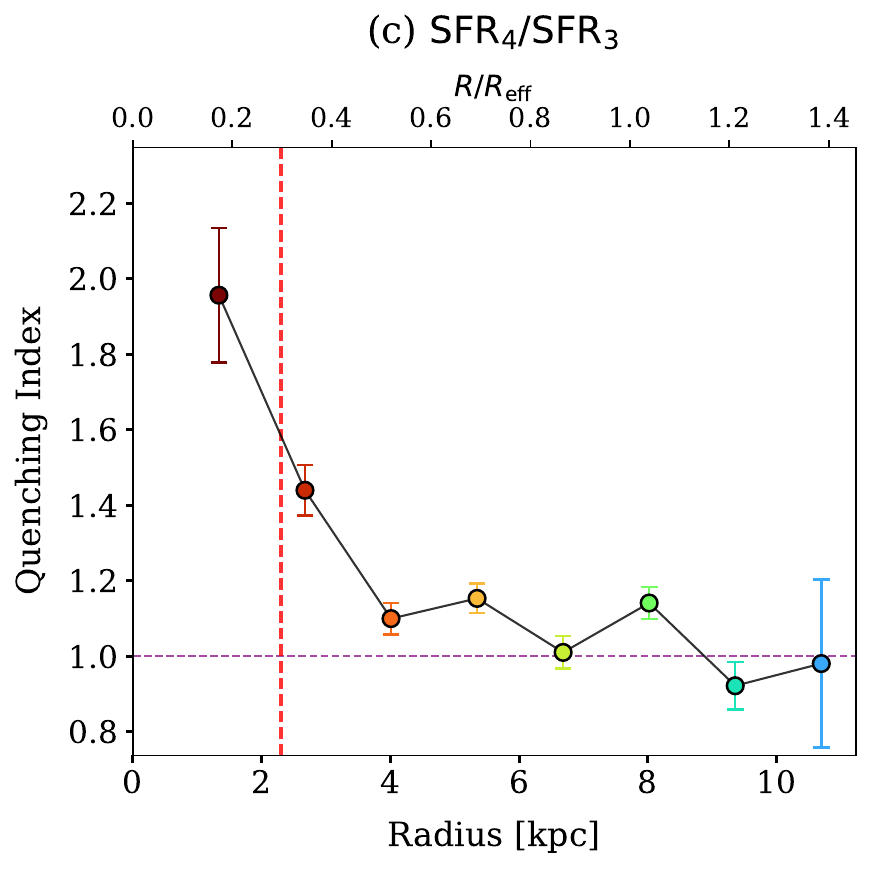}
    \label{fig:SFR_ratio_3}
  }

  \caption{Quenching Index in NGC~6946. Ratios $>1$ indicate quenching, while ratios $<1$ indicate enhancement relative to the previous epoch. Panel (a) shows global quenching during bin~1, panel (b) systematic enhancement in bin~2, and panel (c) declining radial quench. The error bars were estimated using H$\alpha$ fluxes and \texttt{SINOPSIS} uncertainties. The vertical dashed line corresponds to the radius of the secondary bar.}
  \label{fig:SFR_ratios}
\end{figure*}

To better analyse whether NGC 6946 experienced quenching or enhancement of star formation activity, we exploited  the quenching index as defined in \cite{Perez2023} which is the ratio between SFR in two consecutive age bins ($\mathrm{SFR}_{i+1}/\mathrm{SFR}_{i}$) for each region. This approach quantifies how the SFR changed between consecutive epochs, creating a temporal gradient across the galaxy. Values greater than 1 indicate quenching, while values less than 1 correspond to enhanced star formation relative to the previous epoch. The results are presented in Fig.~\ref{fig:SFR_ratios}.

Fig.~\ref{fig:SFR_ratios}a reveals clear quenching across the entire galaxy in age bin 1. The strongest quenching --by far-- occurred in Region 2 (which fully includes the stellar bar), where the SFR decreases by a factor of $\sim$9. Significant quenching also occurred in Region 1, with a decrease by a factor of $\sim$3. The occurring of an intense episode of star formation in bin 2 is also evident in Fig. \ref{fig:SFR_ratios}b, which shows systematic enhancement within this age bin  (all ratios $<$1). Fig.~\ref{fig:SFR_ratios}c shows the galaxy underwent a quenching phase with a clear radial trend that affected nearly its entire disc. This quenching was more pronounced in the innermost regions, consistent with an inside-out assembly scenario where star formation ceases first in the galactic centre. However, a subsequent reversal of this trend occurred in age bin 2: rather than continuing the radial quenching progression observed in age bin 3, the galaxy experienced its peak SFR—an event particularly intense in Region 2.

Regions 1 and 2 increased their SFR from bin 3, peaked in bin 2, and recently quenched in bin 1, despite maintaining substantial molecular gas reservoirs (Fig. \ref{fig:hi_h2_maps}b).  We study the influence of molecular and total gas in the SFR for NGC 6946 in Section \ref{sec:KS}. The current central concentration of gas and intense star formation activity suggests efficient fuelling mechanisms and favourable conditions for star formation in the galactic centre throughout the galaxy's history. 

To assess the robustness of the recovered SFH against the intrinsic degeneracies of stellar population synthesis, we repeated the spectral fitting five times, each time excluding two SSPs of contiguous ages from the template library. This choice was made because SSPs of neighbouring ages have highly similar spectra and can readily compensate for one another in the fit. This experiment was performed with SSPs of ages from $5\times 10^7$ to $10^9$ years. The resulting $\chi^2$ maps were compared with those of the reference model within the elliptical annuli. To determine the importance of the excluded SSPs within each annulus, we computed a number of metrics such as the distribution of $\Delta\chi^2 = \chi^2_{\rm excluded} - \chi^2_{\rm ref}$, together with its median and relative variation, the fraction of spaxels exhibiting larger $\chi^2$ values in the reduced model, and the Wilcoxon signed-rank test with the corresponding rank-biserial correlation as a measure of the effect size. Among the five experiments, only the exclusion of the SSPs in the $\sim 5\times 10^8$ to $8.1\times 10^8$ year interval produce a pronounced and systematic degradation of the spectral fits. In the innermost annuli, more than 90\% \ of the spaxels exhibit increased $\chi^2$ values, while the effect decreases steadily with galactocentric distance and becomes statistically insignificant in the outermost annulus. The remaining age intervals produce either substantially weaker changes or no significant degradation. While, on the one hand, this analysis demonstrates that the inferred intermediate-age stellar population is not a generic consequence of the adopted SSP library, but is specifically required to reproduce the spectra of the inner galaxy, on the other hand it indicates that the dominant intermediate-age population is most consistently associated with SSPs in the 0.5–0.8 Gyr interval.
\subsection{Gas and dust trends}
\label{ss:gas_dust_trends}
\begin{figure*}
  \centering
  {
    \includegraphics[width=0.48\textwidth]{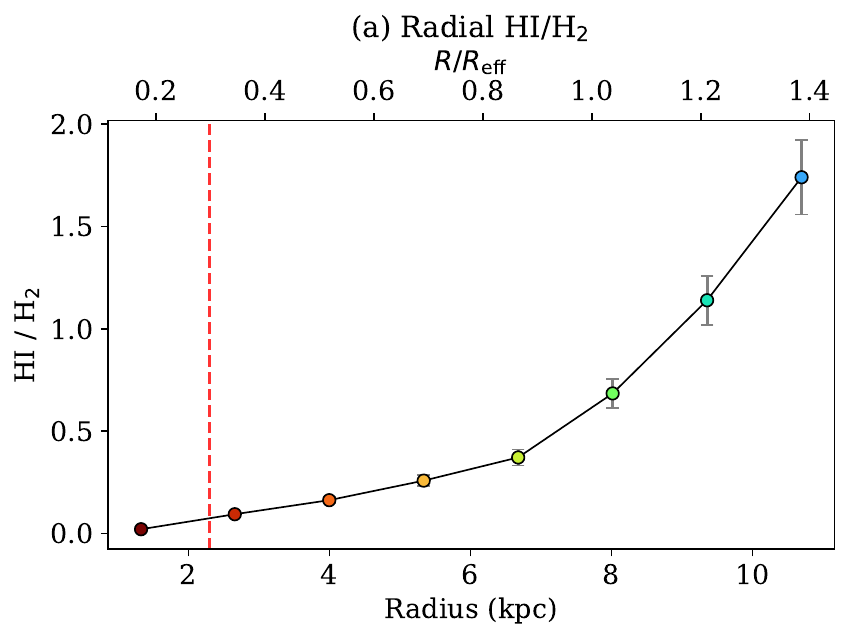}
  }
  \hfill
  {
    \includegraphics[width=0.48\textwidth]{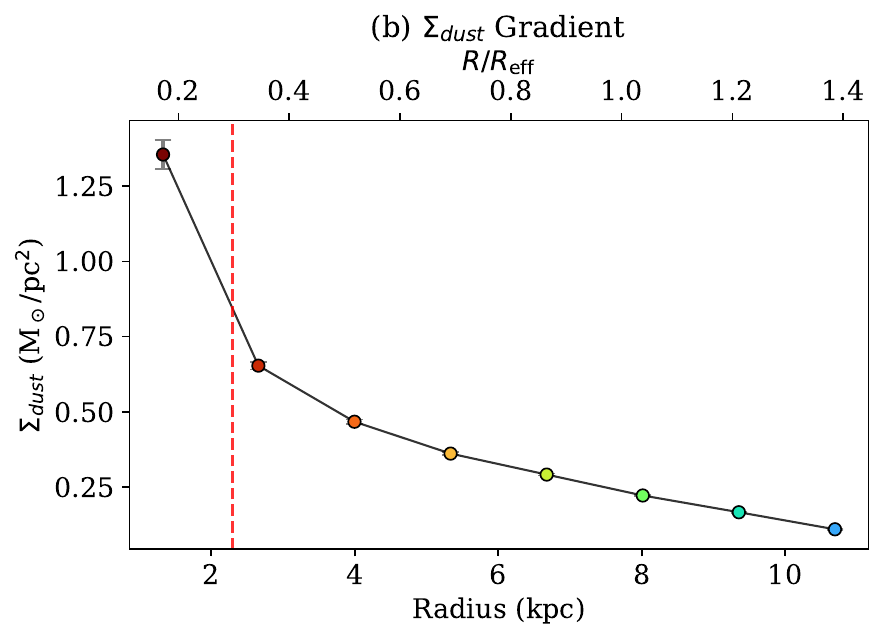}
  }

  {
    \includegraphics[width=0.48\textwidth]{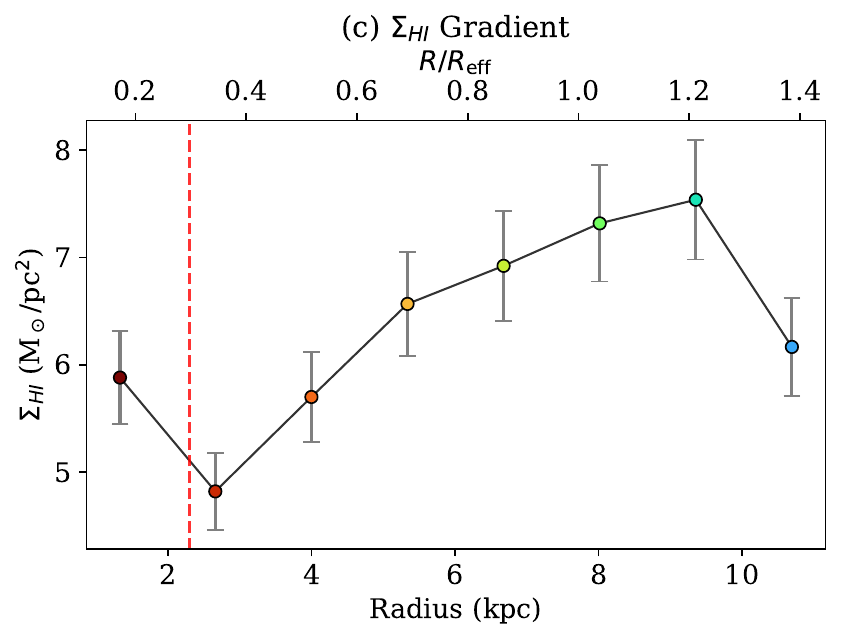}
  }
  \hfill
  {
    \includegraphics[width=0.48\textwidth]{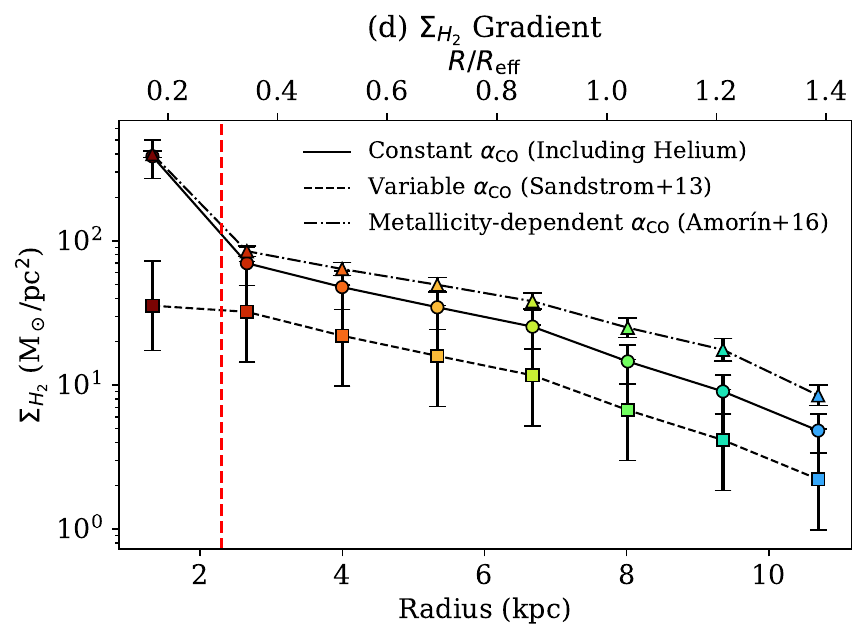}
  }

  \caption{Radial distributions of gas and dust components. Panel~(a) shows the HI-to-H$_2$ mass ratio gradient, which increases steeply beyond $\sim$7~kpc, correlating with reduced star formation activity, while the flatter inner gradient reflects efficient atomic-to-molecular conversion. Panel~(b) shows the dust surface density, which decreases exponentially with radius and peaks in the inner regions. Panel~(c) presents the HI surface density radial profile. Panel~(d) shows the molecular gas surface density ($\Sigma_{\text{H}_2}$), which follows a trend similar to the dust distribution, with a strong central concentration. The solid line corresponds to the constant Galactic $\alpha_{\rm CO}$ prescription (\citealt{Bolatto2013}), the dashed line to the spatially varying conversion factor of \citet{Sandstrom2013}, and the dash-dotted line to the metallicity-dependent prescription of \citet{Amorin2016}. The vertical dashed line marks the radius of the secondary bar.}
  \label{fig:gas_dust_panels}
\end{figure*}

\begin{figure}
    \centering
    \includegraphics[width=0.7\linewidth]{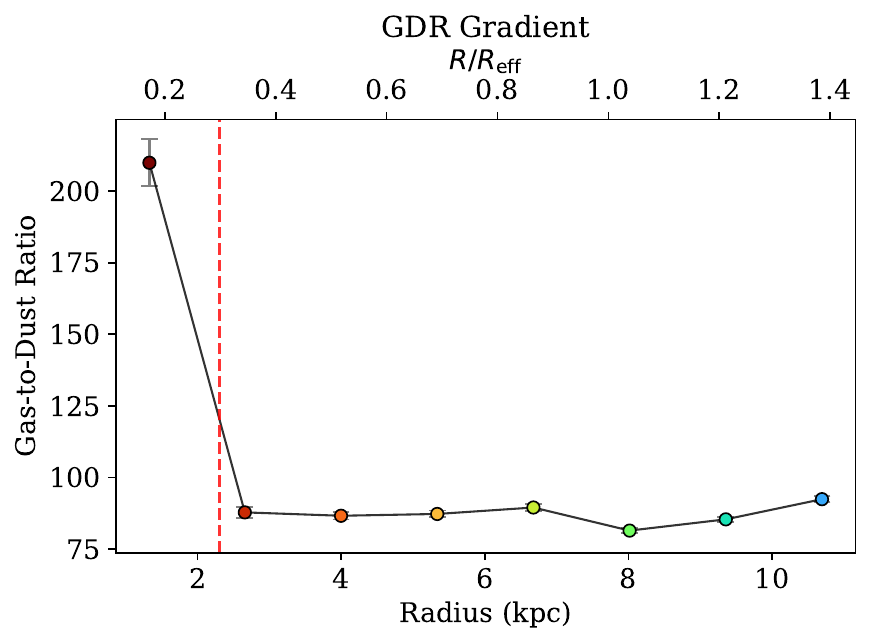}
    \caption{Radial gradient of GDR ratio, showing an overdensity of molecular gas relative to dust in the innermost region, with the ratio increasing towards the outer disc.}
    \label{fig:GDR_gradient}
\end{figure}

The radial distributions of HI, H$_2$, and their ratios across the eight regions are quantified in Table~\ref{tab:gas_radial}. The radial HI/H$_2$ gradient is shown in Fig.~\ref{fig:gas_dust_panels}a. The ratio increases steeply beyond region 5 ($\sim$7 kpc), coincident with shallower $\Sigma_\text{SFR}$ gradients in age bins 1 and 2 (Fig. \ref{fig:SSFR_R}). Inside of a radius of $\sim 7$ kpc, where $\Sigma_\text{SFR}$ remains elevated across all age bins, the HI/H$_2$ gradient is flatter than in external regions. The GDR gradient is presented in Fig. \ref{fig:GDR_gradient}.
\begin{table*}
\centering
\caption{Radial distribution of gas and dust surface densities of NGC 6946.}
\label{tab:gas_radial}
\begin{tabular}{ccccccc}
\hline
Region & $\Sigma_{\text{HI}}$ (M$_\odot$/pc$^2$) & $\Sigma_{\text{H}_2}$ (M$_\odot$/pc$^2$) & HI/H$_2$ & $\Sigma_{\text{dust}}$ (M$_\odot$/pc$^2$) & Gas/Dust \\
\hline
1 & $5.95 \pm 0.02$ & $284 \pm 4.4$ & $0.021 \pm 0.000$ & $1.356 \pm 0.049$ & $210 \pm 8$ \\
2 & $4.84 \pm 0.01$ & $51.4 \pm 0.4$ & $0.094 \pm 0.001$ & $0.653 \pm 0.013$ & $88 \pm 2$ \\
3 & $5.68 \pm 0.01$ & $35.1 \pm 0.2$ & $0.162 \pm 0.001$ & $0.467 \pm 0.007$ & $87 \pm 1$ \\
4 & $6.56 \pm 0.01$ & $25.4 \pm 0.1$ & $0.258 \pm 0.001$ & $0.361 \pm 0.005$ & $87 \pm 1$ \\
5 & $6.90 \pm 0.01$ & $18.6 \pm 0.1$ & $0.370 \pm 0.002$ & $0.291 \pm 0.003$ & $90 \pm 1$ \\
6 & $7.32 \pm 0.01$ & $10.7 \pm 0.1$ & $0.684 \pm 0.004$ & $0.221 \pm 0.002$ & $81 \pm 1$ \\
7 & $7.60 \pm 0.01$ & $6.63 \pm 0.03$ & $1.149 \pm 0.006$ & $0.166 \pm 0.002$ & $85 \pm 1$ \\
8 & $6.13 \pm 0.01$ & $3.55 \pm 0.02$ & $1.731 \pm 0.011$ & $0.109 \pm 0.001$ & $92 \pm 1$ \\
\hline
\end{tabular}
\end{table*}

\begin{figure*}
    \centering
    \includegraphics[width=\textwidth]{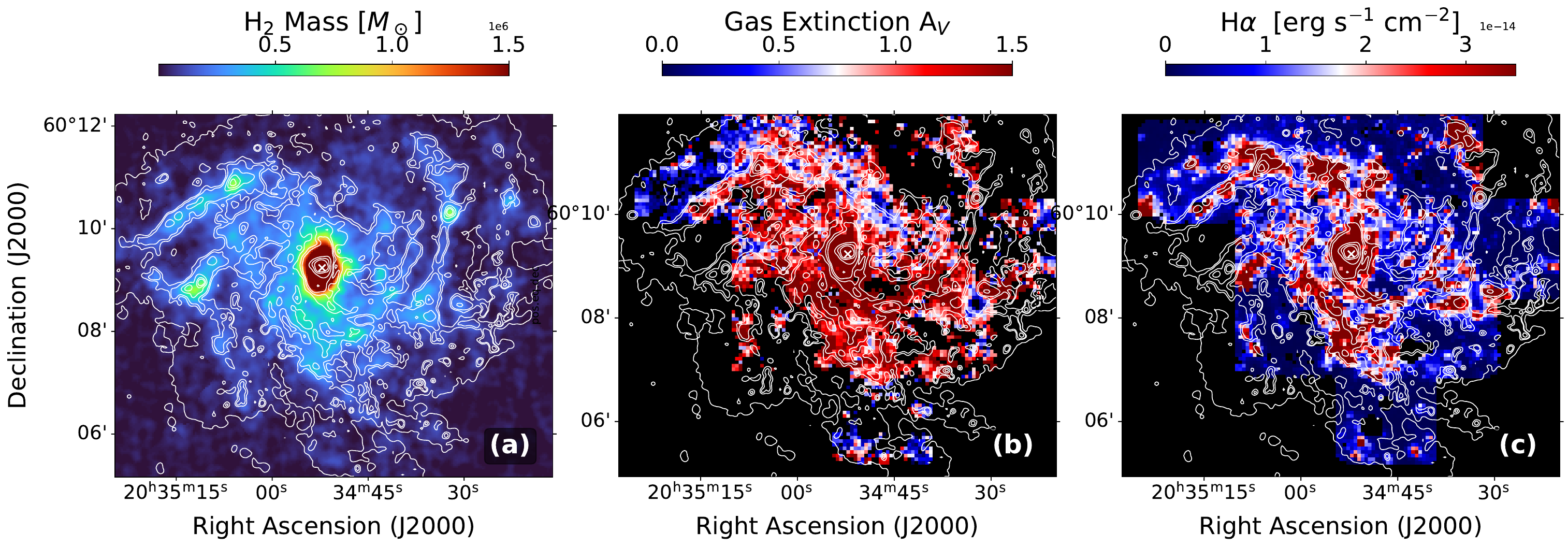}
    \caption{Spatial distribution of gas, dust extinction, and star formation tracers in NGC 6946 with dust mass contours overlaid (white curves). The galaxy centre is marked with a white $\times$ symbol  \textit{Left:} H$_2$ mass map showing molecular gas distribution across the galactic disc. \textit{Centre:} Gas extinction ($A_V$) map derived from the Balmer decrement, indicating dust attenuation. \textit{Right:} H$\alpha$ emission map tracing recent star formation activity.}
    \label{fig:contour_overview}
\end{figure*}

To investigate the locations of molecular gas and dust overdensities, we overlaid contours from the highest resolution dust map (Fig. \ref{fig:dust2pt}). These dust mass contours ($\sim 2, 3, 5, 7, 8, 16, 24,  30\times \sigma_{RMS}$) are plotted on maps of molecular gas mass, Balmer-derived extinction, $A_V$, and H$\alpha$ emission. Fig. \ref{fig:contour_overview} shows a strong spatial correlation between dust and molecular gas. The dust contours closely trace the central overdensity and the accumulations within the spiral arms. The arms are well delineated, demonstrating that regions of molecular gas accumulation consistently correspond to dust overdensities. Dust contours also align with regions of elevated extinction, which is physically expected, despite the fact that the methods and observations to derive both maps are completely different. Dust extinction peaks in the central region, coinciding with the highest dust surface density. Furthermore, the overdensities of dust and molecular gas also coincide with H$\alpha$ emission, again in the central region and along the spiral arms. The consistent alignment of dust contours with molecular gas peaks, high extinction regions, and H$\alpha$ emission highlights the spatial correlation between dust, molecular gas reservoirs, and ongoing star formation throughout the galaxy.

To evaluate the sensitivity of our results to the adopted CO-to-H$_2$
 conversion factor, we recomputed the molecular gas surface density ($\Sigma_{\text{H}_2}$) using two alternative prescriptions in addition to the standard Galactic value adopted throughout this work. First, we adopted the metallicity-dependent conversion factor proposed by \citet{Amorin2016}, using the mean oxygen abundance measured in each elliptical region. Second, we considered the spatially varying conversion factor reported by \citet{Sandstrom2013} calculated for this galaxy, adopting their central value of $\text{log}(\alpha
_{\text{CO}})=-0.40\pm0.31$ for Region 1 and their mean disc value for the remaining regions. Fig. \ref{fig:gas_dust_panels}d compares the resulting radial $\Sigma_{\text{H}_2}$ profiles. For this comparison only, we account for the contribution of helium by applying a factor of 1.36 to the values derived using a constant $\alpha_{\text{CO}}$, consistent with the \cite{Sandstrom2013} and \cite{Amorin2016} prescriptions, which already include helium. Our reported molecular gas masses in other parts of the text do not include the contribution from helium.

Although the absolute molecular gas surface densities change, the qualitative radial trend remains congruent in all three cases, with the central region consistently exhibiting the highest molecular gas concentration. The metallicity-dependent prescription predicts an even steeper central enhancement, whereas the \citet{Sandstrom2013} conversion factor produces a considerably flatter profile owing to the low $\alpha_{\text{CO}}$ inferred for the central region. The exceptionally low central conversion factor inferred by \citet{Sandstrom2013} remains a relatively open issue for NGC 6946. As discussed extensively in their work, different observational techniques lead to substantially different estimates of $\alpha_{\rm CO}$ in the centre. Dust-based and multi-line modelling favour values significantly below the Galactic conversion factor, whereas virial analyses of resolved GMCs \citep{Donovan2013} recover values much closer to the Galactic calibration. Furthermore,  adopting a constant Galactic $\alpha_{\rm CO}$ for NGC 6946 is consistent with recent multi-galaxy resolved ISM studies. 
\cite{Tailor2026}, studying a sample of nearby spirals, adopt the metallicity-dependent conversion factor of \cite{Amorin2016} for CO-bright galaxies, including NGC 6946. 
Due to the near-solar gas-phase metallicity of NGC 6946 (12+log(O/H) $\sim 8.58$), the \cite{Amorin2016} prescription yields CO-to-H$_2$ values and radial $\Sigma_{\text{H}_2}$ profiles 
that are very similar to the Galactic \cite{Bolatto2013} benchmark across most of the disc (see also Fig. \ref{fig:gas_dust_panels}d), confirming that the overall radial trends and central gas concentration remain
robust regardless of the specific $Z-$dependence assumed.

Having said that, we retain the constant Galactic conversion factor adopted by \citet{Bolatto2013} for the remainder of this work, which is also consistent with the role of the dust concentration discussed here (see Section \ref{sec:discussion_atomictomolecular}). Most importantly, our conclusions remain consistent with any conversion factor: all three prescriptions preserve the central molecular gas concentration, even though to different degrees, and are therefore consistent with the interpretation of enhanced gas accumulation towards the inner kiloparsecs.

These results stress that the central region is particularly distinct from the disc. It exhibits a continuous burst of star formation throughout the history of NGC 6946, co-spatial with the galaxy's largest reservoir of molecular gas and highest concentration of dust. While SFR, dust, and molecular gas are co-spatial along the spiral arms, this relationship is even more pronounced in the central region. This is particularly interesting as our results indicate that star formation has been ongoing there persistently, suggesting mechanisms such as a stellar bar may be efficiently and persistently funnelling material into this area.

\section{The Kennicutt-Schmidt relation}
\label{sec:KS}

To investigate the relationship between star formation and gas content, we examined the KS relation \citep{Schmidt1959, Kennicutt1998} in NGC 6946 using combined  optical and radio/sub-millimetre data. The KS law quantifies star formation efficiency by relating gas surface density to $\Sigma_{\text{SFR}}$: 
$\Sigma_{\text{SFR}} \propto \Sigma_{\text{gas}}^n$,
where $n$ is the power law index of the relation. Values for $n$ in the star formation law have been extensively studied with a variety of observations in different galaxies, from $\sim 0.8-2$  \citep{Schmidt1959,Kennicutt1998,Bigiel2008,Casasola2022}. This fundamental relation describes how efficiently galaxies convert their gas reservoirs into stars. In order to derive the spatially resolved version of this relation, we divided the optical disc ($R_{25}$) into a grid of regular hexagons, ensuring complete spatial coverage. This approach was adopted to overcome the intrinsic limitations of the H$\alpha$ map, which exhibits irregular discontinuities (see Fig.~\ref{fig:contour_overview}b). For hexagonal bins containing H$\alpha$ 'holes'—regions with missing or non-detected pixels—a uniform surface density was assumed, derived from the detected pixels within those holes. The hexagon extension (vertex to vertex distance) was set to 16$''$ , i.e. $\sim$600 pc physical scale, slightly exceeding the worst resolution among our datasets (CO data). We performed a fit to the KS relation, considering HI alone, H$_{2}$ alone, and the sum of the two gas components. The molecular and total-gas KS relations show moderate correlations, whereas the HI relation exhibits virtually no correlation ($R^2=0.07$), in agreement that atomic gas alone is not directly associated with ongoing star formation, consistent with previous studies \citep{Bigiel2010,Bigiel2011,Schruba2011}. The slope obtained for the molecular gas relation ($n=0.96\pm0.04$) agrees well with previous determinations for NGC~6946 despite methodological differences. In particular, \citet{Casasola2022} employ FUV+22\,$\mu$m star formation tracers at $\sim$1 kpc resolution, whereas our analysis uses extinction-corrected H$\alpha$ measurements at $\sim$600 pc resolution.

\begin{figure}
  \centering
  {
    \includegraphics[width=0.4\textwidth]{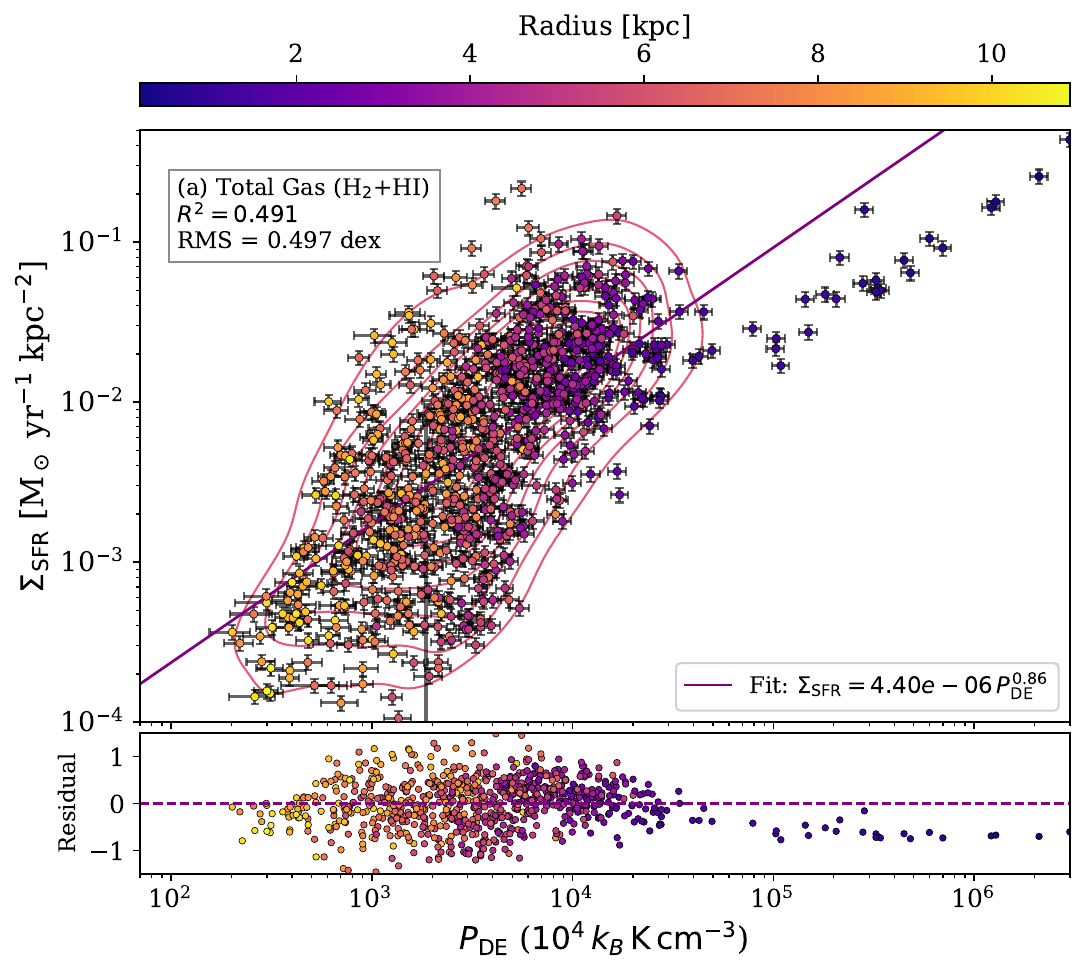}
    \label{fig:KS_H2}
  }
  {
    \includegraphics[width=0.4\textwidth]{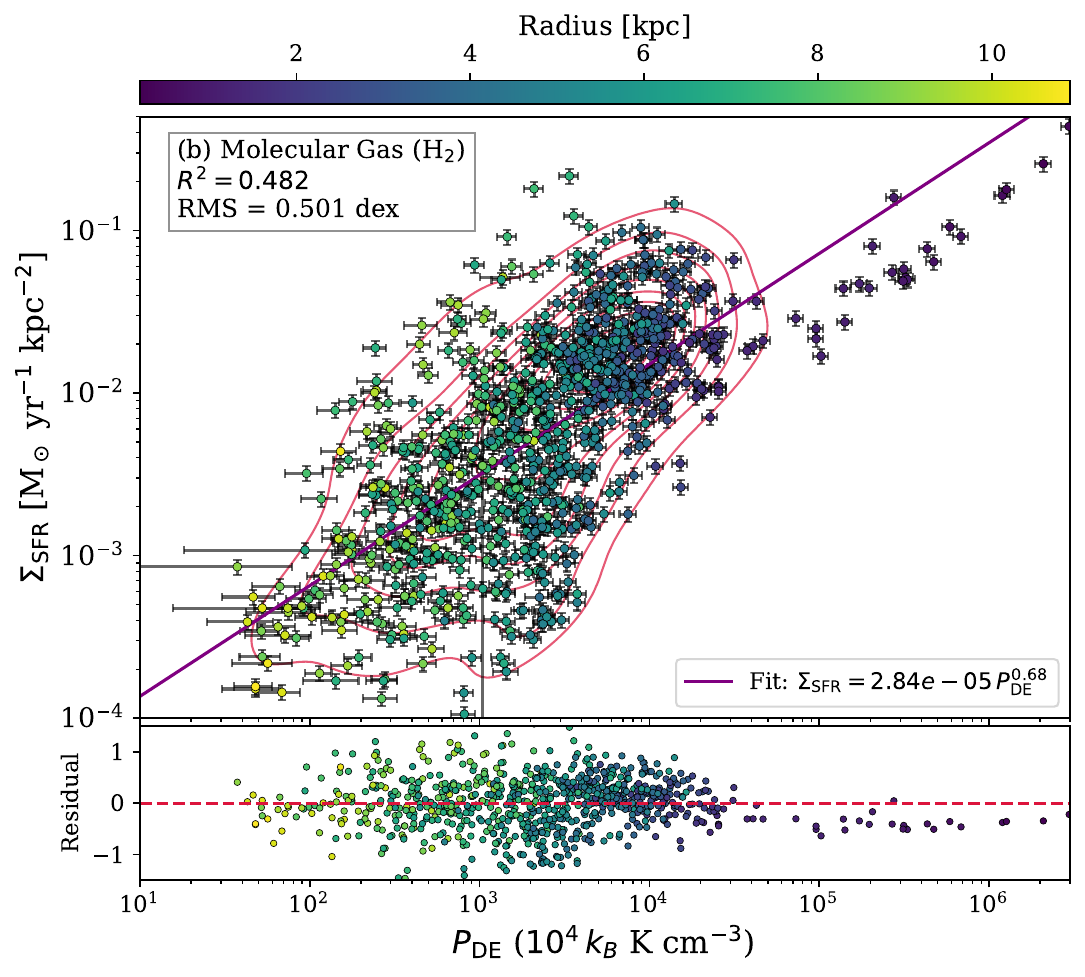}
    \label{fig:KS_total}
  }
\caption{Pressure-regulated star formation relations for NGC~6946. The upper panel shows the conventional total-gas ($\mathrm{H_2+HI}$) relation, while the lower panel shows the H$_2$-only pressure proxy. Points are colour-coded by galactocentric radius, and the lower panels show the residuals. The dynamical equilibrium pressure, $P_{\rm DE}$, is expressed in units of $10^4\,k_{\rm B}\,\mathrm{K\,cm^{-3}}$.}
  \label{fig:KS_relations}
\end{figure}

The scatter in the resolved KS relation may arise from a combination of observational uncertainties and variations in the local physical conditions (e.g. local gravitational potential, ISM pressure, gas velocity dispersion) regulating star formation. While uncertainties in the determination of the SFR and gas surface density contribute to the observed dispersion, departures from the canonical relation may also reflect the influence of additional physical processes that are not captured by a simple dependence on the gas surface density alone. Moreover, as noted in Section \ref{sec:ionized_gas}, optical recombination lines such as H$\alpha$ tend to underestimate the true SFR in heavily dust-obscured environments, particularly in the innermost kpc of NGC~6946. Accounting for this deep obscuration would shift the central spatial bins towards higher $\Sigma_{\mathrm{SFR}}$ at a given gas surface density, indicating that part of the dispersion in the resolved star formation law is driven by heavy central optical attenuation rather than intrinsic variations in star formation efficiency alone. To explore this possibility, we examined two alternative formulations of the star formation law that explicitly incorporate the effects of the local gravitational potential and the dynamical state of the ISM. Following \citet{Shi2018} as a first-order approach, we fitted the extended KS law including the stellar mass surface density as an additional regulator of star formation: $\Sigma_{\rm SFR}
\propto
\left(
\Sigma_{*}^{0.5}
\Sigma_{\rm gas}^{1.09}
\right)^{n}$. The \cite{Shi2018} is a convenient formulation  for a more complex dependence of $\Sigma_{\text{SFR}}$ on $\Sigma_{*}$ \citep{Dib2017}. Second, we evaluated the relation between the SFR surface density and the dynamical equilibrium pressure: $P_{\rm DE}
=
\frac{\pi G}{2}\Sigma_{\rm gas}^{2}
+
\Sigma_{\rm gas}\,
\sigma_{\rm gas,z}\,
\sqrt{2G\rho_*},$ following the formalism of \cite{Ostriker2010} and \cite{Sun2023}, where $\rho_*=\Sigma_*/(2h_*)$. We adopted a stellar scale height of $h_*\sim450$ pc \citep{Kregel2002} and a mean gas velocity dispersion of $\sigma_{\rm gas,z}\sim30$ km\,s$^{-1}$ within the central 1.3 kpc and 10 km\,s$^{-1}$ elsewhere \citep{Romeo2015}. For this analysis, we reproduced the conventional total-gas dynamical-equilibrium pressure using $\Sigma_{\rm gas}=\Sigma_{\rm H_2}+\Sigma_{\rm HI}$ (PDE total; upper panel in Fig. \ref{fig:KS_relations}), and we also considered an H$_2$-only pressure proxy using $\Sigma_{\rm gas}=\Sigma_{\rm H_2}$ (PDE H$_2$; lower panel in Fig. \ref{fig:KS_relations}). The same $\sigma_{\rm gas,z}$ prescription is adopted in both cases. These formulations allow us to assess whether incorporating the local gravitational field and ISM pressure accounts for part of the observed scatter in the resolved KS relation.

Both alternative prescriptions yield only modest improvements over the classical KS relation. For molecular gas, the coefficient of determination increases from $R^2=0.45$ to $0.49$ using the extended Schmidt law and to $0.48$ using the pressure-regulated prescription. Similarly, for the total gas relation, $R^2$ increases from $0.46$ to $0.52$ and $0.49$, respectively (see  Table ~\ref{tab:KS_fits} and Fig. \ref{fig:KS_relations}). These results suggest that the local stellar potential contributes to regulating star formation, as including the stellar mass surface density systematically reduces the scatter in the resolved relation. The improvement obtained with the pressure-regulated prescription is comparable but should be interpreted more cautiously, since the inferred dynamical equilibrium pressure depends on assumptions regarding the stellar scale height and the gas velocity dispersion. Overall, the limited reduction in scatter indicates that neither formulation, which are simplified versions of more complex dependencies, fully accounts for the observed dispersion, implying that additional physical processes and/or observational uncertainties continue to contribute to the scatter in the resolved star formation law. The comparable performance of the two prescriptions suggests that much of the information encoded in the pressure-regulated formulation is already captured by the stellar surface density, while the additional assumptions required to estimate $P_{\rm DE}$ limit any further reduction in scatter. The modest improvement obtained with the extended Schmidt and pressure-regulated relations suggests that the current star formation activity in NGC 6946 cannot be fully described by any single local quantity, including the gas surface density, stellar mass density, or hydrostatic pressure. This indicates that additional processes contribute to the observed dispersion. In the broader context of our results, the residual scatter is consistent with the influence of large-scale dynamical processes associated  with the nested stellar bars. These non-axisymmetric structures drive gas inflows and alter the local dynamical environment in ways that local scaling relations cannot fully capture, supporting a scenario where both local ISM physics and bar-driven secular evolution regulate current star formation in NGC 6946.

One potential source of the observed scatter is the presence of incomplete H$\alpha$ coverage within some hexagonal bins. To evaluate this effect, we computed the H$\alpha$ filling fraction of each hexagon, defined as the percentage of pixels with valid H$\alpha$ emission relative to the total number of pixels enclosed by the hexagon. We find that only $\sim12\%$ of the hexagons have a filling fraction below 70\%. Repeating the KS fits after excluding these low-coverage hexagons yields virtually identical slopes, intercepts, and coefficients of determination. We therefore conclude that the incomplete H$\alpha$ coverage of some hexagons does not contribute significantly to the observed scatter in the KS relations.

\begin{table*}
\centering
\caption{Best-fitting parameters for the star formation laws explored in NGC~6946.}
\label{tab:KS_fits}
\begin{tabular}{llccc}
\hline
KS Relation & Gas tracer & $n$ & $\log A$ & $R^2$\\
\hline
Classical& H$_2$
& $0.96 \pm 0.04$
& $-3.34 \pm 0.01$
& 0.45 \\

Extended & H$_2$
& $0.68 \pm 0.02$
& $-5.85 \pm 0.01$
& 0.49 \\

 $P_{DE}$ & H$_2$
& $0.68 \pm 0.02$
& $-4.55 \pm 0.01$
& 0.48 \\

\hline

Classical  & HI
& $1.16 \pm 0.15$
& $-3.19 \pm 0.03$
& 0.07 \\

\hline

Classical & H$_2$+HI
& $1.42 \pm 0.05$
& $-4.17 \pm 0.01$
& 0.46 \\

Extended & H$_2$+HI
& $0.91 \pm 0.03$
& $-7.25 \pm 0.01$
& 0.52 \\

$P_{DE}$ & H$_2$+HI
& $0.86 \pm 0.03$
& $-5.36 \pm 0.01$
& 0.49 \\

\hline
\end{tabular}
\tablefoot{The exponent, $n$, intercept, $A$, and coefficient of determination ($R^2$)
    are listed for the molecular, atomic, and total gas prescriptions.
}
\end{table*}

Our large-scale results are also consistent with the high-resolution molecular study of the central $\sim1.9$ kpc by \cite{Eibensteiner2022}, who resolve the nucleus and the two nuclear bar ends at 75--150 pc scales. They find pronounced environmental differences between the northern and southern nuclear bar ends, with the southern bar end exhibiting enhanced star formation, a larger dense-gas fraction, and stronger shock tracers, indicating that the molecular ISM responds differently along the bar. Their detection of shocked gas and spatially varying dense-gas properties provides evidence that bar dynamics strongly influence the central molecular medium. Although our analysis focuses on the large-scale stellar bar (at $\sim600$ pc resolution for the KS analysis), both studies consistently indicate that gas surface density alone cannot fully account for the observed star formation activity. Instead, local star formation appears to be regulated by the combined effects of gas accumulation, shocks, and dynamical processes associated with the nested bar system.

So far, our results show NGC 6946 experienced intense star formation peaking $\sim 600$ Myr ago, with the central region (1.33 kpc) maintaining consistently elevated (although recently decreasing) SFR. However, based on the coefficient of determination $R^2$, the KS analysis reveals that molecular gas reservoirs alone cannot fully explain this enhanced central star formation. We now investigate additional physical mechanisms that may drive the observed SFR burst, particularly in the central region.

\section{Discussion}
\label{sec:discussion}
At this stage, we have characterised the main stellar population properties of NGC 6946, finding evidence of a sustained SFR, significant dust extinction, relatively young populations and  $\sim40\%$ of the total stellar mass  concentrated within the central $\sim2.6$ kpc. This region also hosts the highest concentrations of molecular gas and dust in the galaxy and shows a marked deficiency of HI, particularly within the radial extent of the secondary bar.
By causing strong gravitational torques, bars redistribute angular momentum between the disc components: while the stellar component supports the bar through elongated orbits, the collisional gas responds by forming shocks along the bar. These shocks trigger star formation and cause the gas to lose angular momentum, driving it inwards. This process often leads to enhanced central concentrations of molecular gas, nuclear star formation, and the growth of central structures such as nuclear rings or pseudobulges, playing a primary role in the secular evolution of disc galaxies \citep{Kormendy2004}. For these reasons, the properties of the bar(s) in NGC 6946 are expected to have a direct impact on its central gas content and star formation activity.

\begin{figure}
    \centering
    \includegraphics[width=0.8\linewidth]{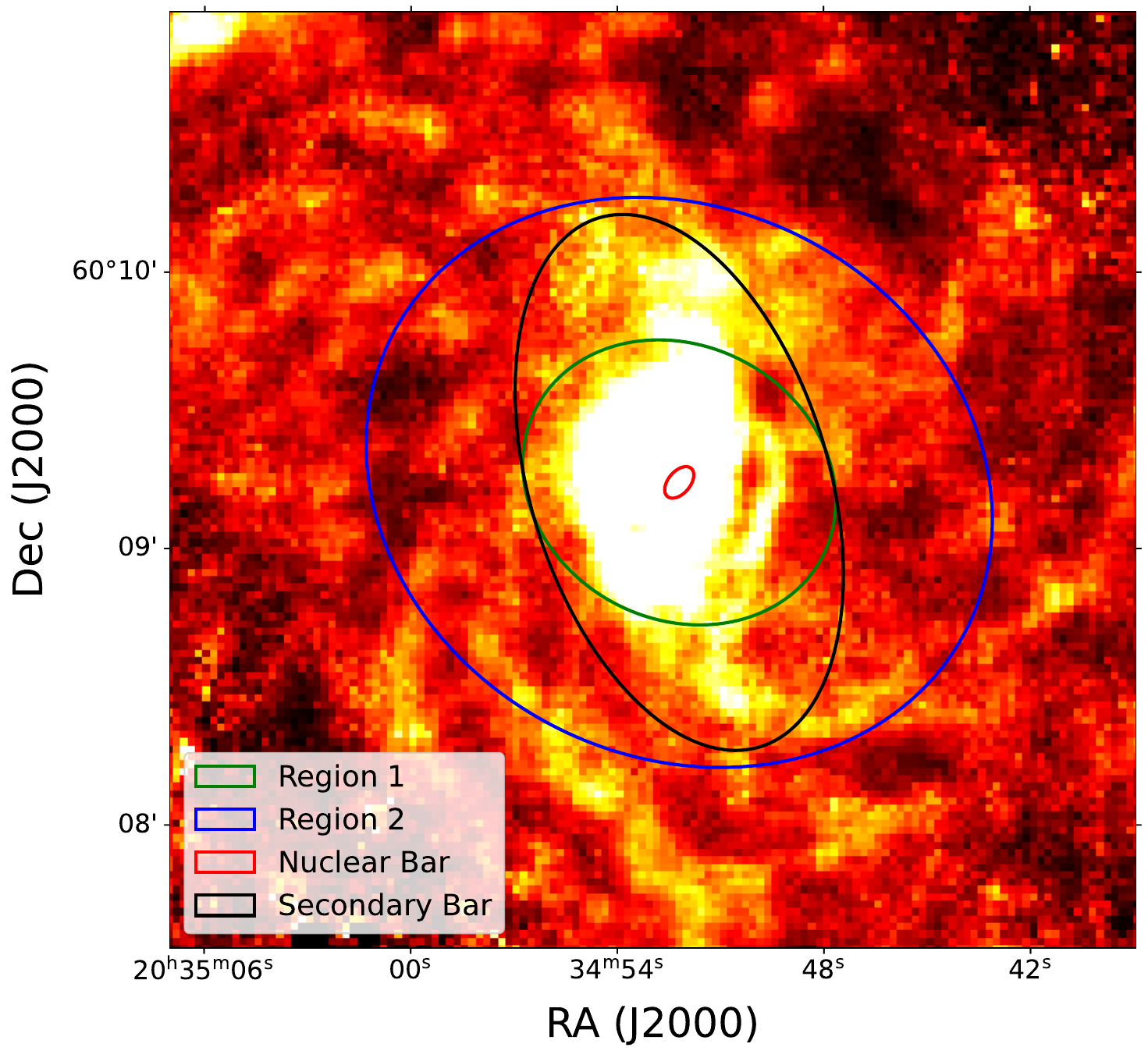}
    \caption{Secondary (black) and nuclear (red) bars of NGC 6946 along with our radial Regions 1 and 2. The background image is the same as in Fig. \ref{fig:dust2pt}.}
    \label{fig: bars}
\end{figure}

Using NIR K-band observations at an effective resolution of $1.7\arcsec$ and CO(1–0) emission data at $4.0\arcsec$ resolution from the Berkeley-Illinois-Maryland Array (BIMA) \citep{Snyder1994}, \cite{Regan1995} found that while a stellar bar of $60''$ had been inferred from previous CO observations, the only non-axisymmetric features at that scale were spiral arms. However, they also identified a bar-like feature in the K-band isophotes with a length of $120''$ and a P.A. of $30^\circ$. They suggested that this structure could be attributed to the mixing and superposition of NGC 6946's (at least) four spiral arms, as the gas behaviour did not match expectations for a barred system. Nevertheless, they could not categorically rule out the presence of a stellar bar.  In contrast, \cite{Elmegreen1998} identified a small nuclear stellar bar using higher-resolution J and K-band observations. The nuclear bar has a size of $\sim 8''$, P.A. of $140^\circ$, and an ellipticity of 0.40. It exhibits a flat light distribution and a strong contrast of 1 magnitude between the bar and inter-bar regions, indicating a high stellar density -- characteristics typical of strong, large-scale bars. According to \citet{Elmegreen1998,Schinnerer2007, Romeo2015} this nuclear bar is likely fuelling the nuclear starburst in NGC 6946  at this scale.\\
\cite{Elmegreen1998} also found a larger stellar oval distortion of length $\sim 70''$ with a P.A. of $\sim 35^\circ$ and an ellipticity of 0.15. They associated this with a bar-within-bar geometry, typical of nested stellar structures. Furthermore, they found that the innermost CO distribution of \cite{Regan1995} perfectly matches the orientation of the nuclear bar, and that the rigid-body rotation of the gas extends to $3$–$4''$, precisely the extent of the nuclear bar.\\ 
Although the nuclear bar is small, there is compelling evidence of its strong dynamical influence in the central region of NGC~6946. \citet{Schinnerer2007} conducted high angular resolution ($\sim 1''$ and $0.6\arcsec$) millimetre interferometric observations of the $^{12}$CO(1--0) and $^{12}$CO(2--1) line emission in the central 300 pc using the IRAM Plateau de Bure Interferometer (PdBI). At this sub-arcsecond resolution, they found the molecular gas distributed in an $S$-shaped spiral structure, a morphology they attribute to the gas responding to the non-axisymmetric potential of a small-scale inner stellar bar. Their work further suggests that the innermost $\sim 100$~pc of NGC 6946 is kinematically distinct with respect to the outer regions, and indicate that molecular gas properties in galactic nuclei can be highly variable over time, critically depending on the nuclear SFH and the precise shape of the stellar potential at the epoch of observation.\\
The larger oval distortion identified by \citet{Regan1995} was confirmed by \citet{Menendez2007} using near-infrared $J$, $H$, and $K_s$ images from 2MASS with an angular resolution of $\sim 2$--$3''$, who measured a P.A. of $17^\circ$. They performed an ellipse-fitting technique employing the  full two-dimensional light distribution, requiring a combined ellipticity and P.A. signature to identify the bar. They applied this automated analysis to over 150 spiral galaxies, including NGC~6946, by requiring a combined signature in both ellipticity and P.A. for bars with $\varepsilon > 0.2$. Given the robustness of this method, we adopt the bar parameters for NGC~6946 derived from their study. The parameters of this secondary stellar bar are presented in Table~\ref{tab:structures} along with the stellar oval structures that we will use in the discussion that follows. The studies summarised above, among others, highlight the challenge in characterising the multiple inner structures present in NGC~6946. We show the positions of the secondary and nuclear bars relative to our Regions 1 and 2 in Fig. \ref{fig: bars}. Both bars are fully contained within these regions, allowing us to quantitatively assess their influence using our results. 

\begin{table}
    \centering
    \caption{Structural parameters of the stellar bars and oval distortion (SOD) in NGC~6946.}
    \begin{tabular}{lcccc}
        \hline
        Structure & $a$ (arcsec) & $a$ (kpc) & P.A. ($^\circ$) & $\varepsilon$ \\
        \hline
        Secondary Bar \, $^a$ & 60 & 2.28 & 17 & 0.46 \\
        SOD\,$^b$ & 35 & 1.33 & 35 & 0.15 \\
        Nuclear Bar\,$^c$ & 4 & 0.15 & 140 & 0.40 \\
        \hline
    \end{tabular}
    \tablefoot{$a$, P.A., and $\varepsilon$ denote the semi-major axis,
    position angle, and ellipticity, respectively, of each structure.}
    \tablebib{
    $^a$~\citet{Menendez2007}, \citet{Regan1995};
    $^b$~\citet{Elmegreen1998};
    $^c$~\citet{Elmegreen1998}, \citet{Schinnerer2007}, \citet{Fathi2007}.
}
    \label{tab:structures}
\end{table}


\subsection{Bar-induced central properties}

As established in numerous studies, the non-axisymmetric potential of a stellar bar induces large-scale streaming motions in both the stellar and gaseous components of the galaxy (e.g. \citealt{Piner1995}). The collisional and dissipative nature of the gas causes it to lose angular momentum, flowing inwards along the bar's dust lanes and becoming concentrated on the leading edges  of the bar \citep{Athanassoula1992}. This process leads to significant dynamical-driven changes, most notably,  accumulation of gas and dust in the central kiloparsec. This result was quantitatively demonstrated by \citet{Sakamoto1999}, who found that barred spirals exhibit higher molecular gas concentrations in their central regions, with CO-derived gas masses reaching $M_{\text{gas}} \sim 10^8$--$10^9~M_{\odot}$ within $R \sim 0.5$ kpc, using a sample of 20 nearby spiral galaxies ($D<37$ Mpc), and spanning all late-type subtypes (among which NGC 6946). The central molecular gas mass of $\sim 6 \times 10^8~M_{\odot}$ within $\sim 0.5$ kpc supports a scenario where the stellar bars in NGC~6946 are at least partially responsible for the central accumulation of molecular gas. This is consistent with \citet{Kuno2007}, who compared 40 nearby barred and non-barred spirals (including NGC 6946) and found that barred galaxies exhibit significantly higher molecular gas concentrations within radii comparable to the bar length. A recent study that confirms this trend was performed by \citet{Sanchez2026} who find that bars increase gas concentration in central parts, leading to a $L(H\alpha)$ and dust content enhancement, in agreement with our results (Fig.\ref{fig:contour_overview}). The bar-driven inflow of molecular gas to the central kiloparsec was further confirmed by \citet{Sheth2005}, who also found that late-type barred spirals are less centrally concentrated than early types. This trend, coupled with the observation of enhanced star formation in early-type bars, suggests higher mass accretion rates in early spiral types, likely related to the longer, stronger, and flatter bars found in such galaxies. 

In this context, NGC~6946 presents an interesting case: the SFH reconstruction we performed (Section \ref{ss: SFH}) indicates that it has formed stars more intensively in its central region than in its disc throughout most of its SFH, something which is in contrast with its late type spiral nature. In addition, this is likely related to the bars-within-bars structure of the galaxy, including a nuclear bar that hosts an even stronger (compared with the one in its surroundings) starburst \citep{Schinnerer2007}. There is also evidence suggesting that the influence of this bar is responsible for the high nuclear molecular gas density observed in this galaxy \citep{Schinnerer2006}. This behaviour is strongly supported by the findings of \citet{Li2023}, who used controlled simulations to show that nested bars increase gas concentration and trigger star formation even more effectively than single bars. On the other hand, due to the fact that the central region remains the major concentration of molecular gas in the galaxy, it is likely that star formation will  continue, even though it is currently in a declining phase, given that the peak of the SFR occurred $\sim 6 \times 10^8$ years ago (Fig.~\ref{fig:SSFR_R}).

Bar-like structures in NGC~6946, aside from the nuclear bar, are not well established, with no general consensus on their properties. This could be due to the secondary bar and/or stellar oval distortion being progressively disrupted, making them difficult to identify. Evidence supporting this hypothesis comes from the 3D N-body simulations of \citet{Athanassoula2005}, which shows that a central mass concentration (CMC) of a few percent of the disc mass can significantly weaken a bar. The level of this weakening depends on the bar type: exponential-profile bars in disc-dominated galaxies can be completely destroyed, while strong, flat-profile bars only experience a reduction in strength. This weakening occurs from both the inner and outer parts of the bar simultaneously as the CMC grows \citep{Roberts1979, Norman1996, Sellwood1999}. Our results align with these simulations. We find that the bulk of the stellar mass in NGC~6946 is concentrated in the innermost regions (Fig.~\ref{fig:stellar_mass}), with approximately 40\% of the total disc stellar mass enclosed within a radius of $\sim 2.6$ kpc—a region that encompasses the secondary bar, the stellar oval distortion, and the nuclear bar. Furthermore, \citet{Elmegreen1998} report that the stellar oval distortion (contained within the secondary bar) has an exponential profile in the $K$ band, while the nuclear bar displays a flat profile. This would explain why the nuclear bar remains strong compared to the other structures, as the growing CMC affects them differently. Consequently, this would explain our results about the current SFR in the central region, while still active, is declining compared to its peak $\sim 6 \times 10^8$ years ago: the increasing CMC likely weakened the secondary bar, reducing the efficiency of gas inflow and leading to the observed secular decline in SFR. All the former behaviours are well supported by our  results of the SFH presented in Figs. \ref{fig:SSFR_normalized}, \ref{fig:SSFR_R}, \ref{fig:SFR_ratios}. Although the former  scenario about differential weakening of bars is fully consistent with our results, \citet{Li2023} found that the innermost bar in this nested systems (i.e. our nuclear bar)   gradually weakens and dissolves due to gas inflow
over a few billion years. On the other hand, \citet{Erwin2011} found inner bars are either long-lived or easily reformed
after dissolution.

\cite{Fathi2007} performed a two-dimensional kinematic analysis of the H$\alpha$-emitting gas in NGC~6946. Their study allowed them to deduce the location of resonance radii and confirm the presence of a bar-like structure inside the Outer Inner Lindblad Resonance (OILR) of the main oval ($a \sim 4.5'$). This structure, (our secondary bar), rotates at more than twice the pattern speed of the main oval and is located at $49^{+2}_{-4}$ arcsec—a region that falls within our Region 2. Furthermore, they confirmed that a nuclear bar has formed inside the ILR of the secondary bar, effectively placing it within the IILR of the main oval at a radius of $14 \pm 1$ arcsec (within our Region 1). This nested configuration establishes NGC~6946 as a prime example of a bar-within-bar system. More interestingly, according to \cite{Elmegreen1998} the nuclear bar is oriented almost perpendicular to the larger secondary bar, a configuration interpreted as evidence of the presence of the characteristic $x_1$ (aligned with the major axis of the bar) and $x_2$ (perpendicular to the bar) orbital families within the same structure, which are the two principal families of periodic orbits in the disc due to the presence of a bar, and each family appears in a range of radius determined by the resonances. 

We propose the following sequence: the bar-driven dynamics, first funnelled the primordial gas -- initially mostly atomic -- inwards. This concentrated atomic gas was then efficiently converted into molecular gas, a process catalysed by the concurrently accumulating dust. It is this efficient conversion, following the bar(s)-driven inflow, that ultimately resulted in the lack of atomic gas observed in the second most inner region, where the secondary bar extends and has its strongest influence. Due to these dust reservoirs acting as a catalyst for molecular gas formation, there has been a continuously renewed reservoir, providing the raw material for the persistent star formation observed in the inner regions throughout the galaxy's history. A cycle of atomic-to-molecular gas conversion has thus fuelled the central molecular reservoir, sustaining high levels of star formation and resulting in the observed HI deficiency and molecular gas excess in Region 2. The positive HI/H$_2$ radial gradient (Fig. \ref{fig:gas_dust_panels}a) may be indirectly linked to the negative SFR gradient found for age bin 2, with this process being most efficient within the central bar region and declining at larger radii. 

The observed gas accumulation in the central region may be associated with resonance locations (e.g. ILRs), where gas tends to pile up due to orbit crowding. The bars or nested structures may be producing a cascade effect from the outermost structures to the nuclear bar. This is supported by the  findings of \citet{Fathi2007} on the kinematical structure of the innermost regions and by \citet{Romeo2015} who found evidence of molecular gas inflow in the innermost $\sim 1$ kpc of this galaxy, using radial profiles of velocity dispersion and LoS velocity from CO data. These provide  compelling dynamical evidence of the excess of dust and molecular gas at recent times (see Section \ref{sec:discussion_atomictomolecular}), and subsequently the intense H$\alpha$ emission (tracing ionised gas, due to the triggering of SF), and the high dust extinction we observe in the central regions. As shown in our results (Fig.~\ref{fig:contour_overview}), these features display a strong spatial correlation.  

To provide further indirect evidence supporting inward gas flow from Region 2 to Region 1 driven by the secondary bar, we estimated the effective oxygen yield, $Y_{\text{eff}}$, which provides a diagnostic of gas exchange through deviations from the closed-box expectation \citep[e.g.][]{Pagel2009, Matteucci2012, Pilyugin2025}. We computed $Y_{\text{eff}} = Z / \ln(1/\mu)$, where $\mu$ is the gas mass fraction and $Z$ is the oxygen mass fraction derived from O/H. Since emission-line abundances we derived trace only the gas phase, we corrected for oxygen depletion onto dust by applying a +0.12 dex offset appropriate for metal-rich H\,II regions \citep{Peimbert2010}. We obtain $\mu = 0.173$ and $Y_{\text{eff}} \approx 3.8 \times 10^{-3}$ in Region 1, and $\mu = 0.108$ and $Y_{\text{eff}} \approx 2.7 \times 10^{-3}$ in Region 2, consistent with empirical estimates of the true oxygen yield  \citep[$Y_O \sim 0.003$; e.g.][]{Pilyugin2004,Bresolin2004}. The obtained behaviour of $Y_{\text{eff}}$ can be well explained within the framework of radial gas flow into the centre. The radial metallicity gradient in the galaxy is shallow, $\sim 0.02$ dex kpc$^{-1}$ (see Eq.~\ref{eq:OH_grad}). Therefore, radial gas inflow increases significantly the gas mass fraction in the central region, while the decrease in the oxygen abundance remains small. These two effects, namely a significant increase in $\mu$ and a modest decrease in metallicity, result in an enhanced value of $Y_{\text{eff}}$ in Region 1 and at the same time in a decreasing in $Y_{\text{eff}}$ and $\mu$ in Region 2, as observed in our results. These fact strength the conclusion about the gas inflow due to the secondary bar. It should be emphasised that the radial flow of high-metallicity gas into the centre and infall of the pristine results in opposite change in the obtained value of the effective oxygen yield. The infall of the pristine gas mimics the decrease in the oxygen yield while the radial inflow of the oxygen-rich gas mimics the increase in the oxygen yield.

Although we do not directly measure radial gas velocities, several of the observed quantities—including the enhanced central molecular gas concentration, the radial increase in effective oxygen yield, the elevated gas fraction, and the dust properties—are naturally expected as consequences of sustained inward gas transport. These diagnostics therefore provide indirect evidence consistent with gas inflow, rather than a direct kinematic detection. Complementary to this, high-resolution CO observations by \cite{Schinnerer2006, Schinnerer2007} revealed non-circular molecular gas motions and streaming along the nuclear bar towards the nuclear region. \cite{Fathi2007} also confirmed a gas inflow through residual H$\alpha$ velocity fields after subtraction of the rotational component. Our results therefore provide an independent chemical and star-formation-based perspective that is consistent with previous kinematic studies.

For this reason, we favour a scenario in which secular transport driven by the nested bar system is the primary mechanism, while gravitational instability may act in concert by promoting the fragmentation of the accumulated gas once sufficiently high central surface densities are reached. In this picture, the bars continuously transport gas towards the central kiloparsec, where it accumulates over several hundred million years. Once a sufficiently high gas surface density is reached, gravitational fragmentation and star formation become highly efficient, giving rise to episodic bursts of star formation. Hydrodynamical simulations of barred galaxies and nuclear rings predict this type of behaviour, in which the SFR responds to the gas inflow rate and exhibits recurrent bursts separated by relatively quiescent phases \citep{Seo2013, Seo2019}. Although the available observations do not establish a direct causal connection between the nested bars and the star formation episode inferred from the SFH, the temporal coincidence between the two, together with the observed molecular gas streaming motions and small pseudo-bulge, lends support to bar-driven gas inflow as the most plausible origin of the enhanced central star formation. As mentioned in Section~\ref{ss:gas_dust_trends}, NGC~6946 is among the nearby galaxies with the largest systematic uncertainties in the determination of $\alpha_{\rm CO}$. This uncertainty could, in principle, affect
some of our arguments. Further investigation is therefore required to provide a definitive answer and better constrain the physical processes at play in this galaxy.

\subsection{High concentration of molecular gas in the central region}
\label{sec:discussion_atomictomolecular}
Now, we focus more on the nature of the ISM and on the transformation between atomic and molecular gas and dust properties  within the central $\sim 2$ kpc of NGC~6946, to unravel if this excess is exclusively due to the influence of the bar(s) or if there is an additional mechanism enhancing the atomic-to-molecular gas conversion. As demonstrated throughout this work, this region also exhibits the most intense concentration of dust in the galaxy (Fig.\ref{fig:gas_dust_panels}b). It is well established that molecular hydrogen forms efficiently on the surfaces of dust grains, which act as catalysts, and that the presence of H$_2$ is strongly correlated with the presence of dust \citep{Hollenbach1971}. The H$_2$ formation rate in the diffuse ISM has been estimated at $R_{\text{H}_2} \sim 3 \times 10^{-17}$ cm$^3$ s$^{-1}$ and appears to be relatively invariant  in normal conditions (diffuse clouds, \citealt{Gry2002}). However, \citet{Habart2004} conducted a study to determine the H$_2$ formation rate in photodissociation regions (PDRs) and compare it with that of diffuse ISM clouds. They found that the formation rate is highly sensitive to the local environment. In quiet, low irradiated (LI) PDRs exposed to moderate UV radiation , H$_2$ forms at a rate five times higher than in diffuse ISM conditions, while in highly irradiated PDRs (HR), the rate is comparable to that in diffuse clouds. This indicates that the relatively extreme conditions in PDRs do not inhibit molecular gas formation and can even accelerate it in the presence of dust and PAHs \citep{Hollenbach1997,LeBourlot2012,Boschman2015,Wakelam2017}. The authors propose that this enhancement occurs through a mechanism of indirect chemisorption on small dust grains ($<100$ Å), where the grain surface properties allow the formation rate to exceed the standard value. Small grains and PAHs experience rapid temperature fluctuations, spending most of their time at low temperatures ($\sim20$K) despite brief UV-induced heating \citep{Guhathakurta1989}. This behaviour enables efficient H$_2$ formation via the Langmuir-Hinshelwood mechanism and indirect chemisorption, making them dominant contributors to molecular gas formation, particularly in warm environments such as moderate PDRs \citep{Bron2014,Cazaux2004}. In the specific case of NGC~6946, \citet{Eibensteiner2022} identify at least 10 molecular species in the galactic centre, indicating a rich molecular chemistry despite the harsh physical conditions associated with shocks and PDRs. These findings may indicate that the central region provides favourable conditions for sustaining a chemically rich molecular ISM, consistent with our interpretation of long-term gas accumulation, enhanced star formation and the important role of dust as catalyst. Having said that, the fact that H$_2$ is formed in a wide variety of astrophysical environments suggests that both physisorption and chemisorption operate in parallel and both large and small dust grains are  more effective catalysts in distinct environments \citep[diffuse ISM and PDRs, respectively][]{Wakelam2017}. On the other hand, \citet{Arab2012} report that for large grains, temperatures derived from MBB fits and radiative transfer models, range from 13–30K for LI PDRs to 25–70K for HR PDRs. Using the MBB method, we derived temperatures up to $\sim 29$ K in the central region of NGC~6946 (see Fig.~\ref{fig:dust_properties_5pt}), suggesting similar physical conditions as in LI PDRs in the galaxy's centre.

Furthermore, evidence suggests that coagulation of dust, where larger grains are formed from small grains, occurs in dense clouds, where a significant fraction of small grains  coagulate within a period of $\sim10$ Myr, modifying the grain size distribution \citep{Hirashita2009}.  \citet{Arab2012} also note that deeper within PDRs, where the UV field is weaker and gas density higher, small grains likely coagulate onto the surfaces of larger grains. Compared to the diffuse ISM, an increase in the emissivity of large grains suggests coagulation. Our results do not show an increase in $\beta$ in Region 1, but a decrease, which may indicate that the population of small dust grains are not coagulating at all into larger grains, at least not totally. Conversely, the increase in the dust emissivity index $\beta$ observed in Regions 2 and 3 suggests enhanced synthesis or coagulation of larger dust grains in these areas. The subsequent inflow of this processed dust -- along with smaller grains -- into Region 1, likely driven by the secondary stellar bar, provides an additional mechanism for accumulating the galaxy's largest dust reservoir in its innermost region, in addition to that due to star formation. 
This framework -- combining bar-driven dynamics and efficient dust-catalysed molecular gas formation -- may explain two important observations. First, the massive concentration of molecular gas in Region 1 results from the bar-driven inflow of the ISM, primarily atomic gas and dust, from Region 2, evidenced by Region 2's atomic gas depletion (Fig.~\ref{fig:gas_dust_panels}c), and the increase in $\beta$ suggesting larger dust grains coagulation that had been transported to the central region. Second, this inflowing atomic gas is converted into molecular form with high efficiency due to the substantial dust reservoir, a process further enhanced by PDRs, whose presence is expected given the persistent star formation throughout the SFH in this region, a hypothesis supported by the higher dust temperatures we measure there. Some of this dust was likely initially funnelled into the centre by the action of stellar bars but also subsequently processed and enriched by the strong, persistent star formation throughout the galaxy's history.\\
Further supporting the pivotal role of dust in molecular gas concentration is the gas distribution across the different regions. In Region 2, we observe the lowest atomic gas concentration in the galaxy, which is compatible with  atomic gas inflow from Region 2 into Region 1. If the molecular gas concentration depended solely on this inflow, Region 2 would also exhibit a low molecular gas content. However, the opposite is true; after Region 1, Region 2 shows the highest molecular gas and dust concentrations. This pattern strongly suggests that the presence of dust may contribute significantly to the efficient conversion and accumulation of molecular gas, rather than inflow dynamics alone. We present these radial gas and dust measurements in Table~\ref{tab:gas_radial}, including the GDRs. This high atomic-to-molecular gas conversion efficiency could contribute to the formation of the double molecular disc structure reported by \citet{Romeo2015}. The nuclear molecular disc, which extends to a radius closely matching that of our Region 1, is described as having an exponential profile with a scale length of $R_1 \simeq 190$ pc. This radial extent is approximately coincident with that of the nuclear bar.

\subsection{Estimation of the last activity epoch of the secondary bar}
\label{subsec:bar_formation}

In this section, we focus on the secondary stellar bar, as our spatial resolution is insufficient for a robust analysis of the nuclear bar. The reconstructed SFH reveals a significant enhancement of star formation $\sim$0.6 Gyr ago following a period of relative quiescence (see Fig.~\ref{fig:SSFR_R}). Given the finite temporal resolution of non-parametric spectral fitting, this age should be interpreted as the characteristic epoch of enhanced star formation rather than the precise timing of a brief burst event. The coincidence of this event with the presence of nested stellar bars, a centrally concentrated molecular gas reservoir, enhanced effective oxygen yield and gas mass fraction supports a scenario in which bar-driven gas inflow contributed to the accumulation of cold gas in the central region and consequently to the enhanced star formation activity. Although alternative mechanisms cannot be excluded, the observed combination of dynamical, gas and stellar population properties is consistent with expectations from a secular bar-driven evolution.

On the other hand, there is some evidence consistent with a young bar. Theoretical and observational studies indicate that pseudo-bulges (or nuclear discs) in barred galaxies form approximately $\sim 10^8$ years after bar formation \citep{Athanassoula1992b,Athanassoula1992,Emsellem2015,Fragkoudi2016}. This delay arises because the newly formed bar induces shocks in the galactic gas, causing it to lose angular momentum and funnel towards the centre, where it accumulates into a nuclear structure. Consequently, the formation time and relative size (with respect to the whole stellar disc) of the nuclear disc serves as a proxy for the bar's age.  \citet{Kormendy2010} report that NGC~6946 hosts only a tiny pseudo-bulge, comprising less than $\sim 3\%$ of the galaxy's total stellar mass. This small mass is consistent with expectations for a relatively young secondary bar, a scenario supported by models in which nuclear discs form from expanding, bar-built nuclear rings \citep{Bittner2020}. In addition, the recent work of \citet{deSa-Freitas2025} finds a tight correlation ($R^ 2=0.85$) between the stellar mass of the nuclear disc and the age of the bar: older bars host more massive, less star-forming nuclear discs, while younger bars are associated with less massive, more actively star-forming discs, such as NGC 6946.

A young bar is also in agreement with the global star formation state of NGC~6946. \citet{deSa-Freitas2025} find strong anti-correlations between the bar age and both the nuclear disc $\Sigma_{\text{SFR}}$ and the galaxy's offset from the star-forming main sequence ($\Delta$MS). Young bars are associated with positive $\Delta$MS values. NGC 6946 resides well above the main sequence, similar to the population of galaxies hosting comparatively young bars \citep{deSa-Freitas2025}. This is also consistent with the idea that bars contribute to galaxy quenching, particularly along most of the bar region \citep{Masters2012,Schawinski2014,Scaloni2024, Gu2026}. Our analysis shows that quenching is occurring across the entire galaxy following the peak in star formation $\sim 600$ Myr ago (Fig.~\ref{fig:SSFR_R}, Table~\ref{tab:sfr_values}), and this quench is most pronounced in Region 2—the full radial extent of the secondary bar—and Region 1 (Fig.~\ref{fig:SFR_ratios}a), but we also find that at the present time, NGC 6946 is well above the main sequence and the quenching is likely still starting. \citet{deSa-Freitas2025} report a mean bar age of $4.0 \pm 2.0$ Gyr for galaxies with high-SF nuclear discs.
Furthermore, simulations suggest that bars in gas-rich, late-type spirals can be short-lived (1–2 Gyr) due to weakening by central CMC and angular momentum transfer from infalling gas \citep{Bournaud2005}. We also estimate that NGC~6946 assembled $80\%$ of its stellar mass in $\sim 8$ Gyr, less than the median of 10.6 Gyr found for barred galaxies by \citet{Fraser2020}, for a given stellar mass and morphological type. They also find evidence of  the coupling
between an early peak in a galaxy’s SFH and the presence of a bar, which suggest a young secondary bar for NGC 6946 given the SFH peak $\sim 0.6$ Gyr ago. As additional evidence, \cite{Fathi2007} reported an elevated pattern speed -- more than twice that of the main disc -- consistent with the presence of a weak and dynamically young bar. Furthermore, the galaxy is dominated by a massive stellar disc ($\log M_* \sim 10.74$), a condition known to favour the development of bar instabilities \citep{Sellwood2014}.

Three plausible scenarios emerge to explain the reported interplay between ISM and stars: (1) the high CMC observed in the centre (Fig.~\ref{fig:stellar_mass}) was already present before the formation of the secondary bar and prevented it from becoming strong; (2) the CMC was largely built by the bar itself, which subsequently weakened it through feedback; or (3) the bar structure is transient, having undergone cycles of weakening and re-assembly, a possibility that cannot be discarded given the complex, nested nature of the stellar structures in NGC~6946, \citep{Norman1996,Athanassoula2005,Bournaud2005,Jang2023,Gadotti2006} that would be consistent with the CMC built by a former bar episode, which now prevents the current bar from becoming strong. A definitive chronology -- whether the CMC preceded or resulted from the bar -- requires a dedicated methodology for dating stellar bars \citep{deSa-Freitas2025}. 

We note, however, that these empirical relations exhibit significant intrinsic scatter and should therefore be interpreted in a statistical rather than deterministic sense. Although the correlations reported by \citet{deSa-Freitas2025} are strong, the inferred bar ages still span a broad range of values, reflecting the complex nature of bar evolution. In particular, properties such as nuclear disc mass, central star formation rate, and offset from the star-forming main sequence  are influenced by several interconnected processes beyond bar age, including gas accretion history, feedback, angular momentum redistribution, and secular evolution \citep[e.g.][]{Athanassoula2003,Kormendy2004}. Consequently, although NGC 6946 occupies the region of parameter space statistically associated with relatively young bars, these observables alone cannot uniquely determine the age of its secondary bar. Instead, these indicators suggest that a comparatively young secondary
bar provides a plausible interpretation that is consistent with the available evidence. The complete dynamical history is further complicated by the nested nature of these bars.

\section{Summary and conclusions}
\label{sec:conclusions}
This work presents a spatially resolved, comprehensive analysis of the ISM and stellar populations in NGC~6946. We have conducted a panchromatic study, combining optical IFU (Metal-THINGS), infrared (\textit{Herschel}), millimetre (HERACLES), and radio (THINGS) data. Utilising the SED fitting code \texttt{SINOPSIS}, we have derived integrated and spatially resolved stellar populations properties, including stellar mass, LW and MW ages, stellar extinction, and the SFH. These were complemented by maps of the atomic and molecular gas mass, and dust properties  (extinction, mass, temperature, and emissivity index, $\beta$) and Balmer-derived extinction. From our optical data, we also detect the ultraluminous X-ray source NGC-ULX~3 via BPT diagnostic diagrams. The total masses of these components within the $R_{25}$ isophote and the GDR are summarised in Table~\ref{tab:mass_components}. We also performed a tilted-ring analysis, dividing the galactic disc into eight radial zones, with Region~1 being the innermost. Synthesising the results of that analysis alongside evidence from the literature, we draw the following conclusions:

\begin{enumerate}
        \item NGC~6946 has maintained a high SFR in its central $\sim 2.6$ kpc throughout its history. This region experienced a  quenching event approximately 5 Gyr ago (between age bins~4 and~3, see Table \ref{tab:sfr_values}, expected in a downsizing scenario), with a clear radial trend of stronger quenching in the centre (Fig.~\ref{fig:SFR_ratios}c). This intense activity built up roughly 40\% of the galaxy's stellar mass in its central regions. The resulting CMC, which could be assembled by a former secondary bar episode, has likely contributed to a gradual weakening of the secondary bar,  which prevented it from growing strong. Following the global SFR peak $\sim$ 600 Myr ago, which can be interpreted as broadly consistent with a recent episode of enhanced bar-driven gas inflow, the galaxy exhibits a decline in star formation, most pronounced along the extent of the secondary bar (Regions 1 and 2; Fig.~\ref{fig:SFR_ratios}a), consistent with bar-driven quenching. Nevertheless, star formation remains high due to the vast molecular reservoir accumulated at earlier epochs.

    \item Several independent observational properties are consistent with
the presence of a comparatively young secondary bar, although none
individually provides a direct measurement of its age. Taken
together, these diagnostics favour this interpretation, while
alternative evolutionary histories cannot be completely excluded. A young bar is consistent with the galaxy’s stellar mass, $\log(M_\star/M_\odot)=10.74\pm0.01$, which places it just above the characteristic mass range ($\log(M_\star/M_\odot)\sim10$ –10.5) where bar instabilities become more efficient, naturally favouring the formation of relatively weak and possibly young bars. This is further supported by the star-forming pseudo-bulge, the strong burst of star formation inferred from our SED fitting peaking at $\sim 600$ Myr and its position  above the SFMS. The timing of this enhancement is consistent with the proposed evolutionary scenario, although it does not by itself establish a causal connection with the secondary bar. In addition, the presence of a well-developed stellar disc and the fast rotation pattern further support a young bar scenario, as these properties are commonly associated with recently formed bars. The quenching signature that we observe is strongest in the central region of the galaxy that is affected by the bar, yet the overall high SFR indicates that this quenching is in its early stages.

\item This evolutionary framework naturally accounts for the elevated SFR of NGC~6946, placing it above the local SFMS (Fig.~\ref{fig:sfms}). The spatially resolved KS analysis shows that gas surface density alone explains only about half of the variance in the current SFR ($R^2\sim0.45$), while including stellar mass or dynamical equilibrium pressure yields only modest improvements. This suggests that local gas properties alone cannot fully describe the current star formation activity. Instead, the residual scatter is consistent with an additional dynamical process associated with the nested stellar bars, which can sustain the long-term transport of gas towards the central region. This interpretation is supported by the enhanced effective oxygen yield ($Y_{\rm eff}\approx3.8\times10^{-3}$) and gas mass fraction ($\mu=0.173$) measured in the centre, compared to the surrounding disc ($Y_{\rm eff}\approx2.7\times10^{-3}$, $\mu=0.108$). Together with  previously reported molecular gas streaming motions in CO and H$\alpha$, these observations point to the inward redistribution of enriched gas, increasing the central gas fraction while only mildly affecting the metallicity owing to the shallow abundance gradient. This behaviour contrasts with the expectations for pristine gas accretion, which would instead reduce $Y_{\rm eff}$, making bar-driven gas redistribution the more plausible explanation for the enhanced central star formation.
 
\item Rather than being driven by a single bar, the gas inflow in NGC 6946 may arise from a hierarchy of non-axisymmetric structures, including a large-scale oval, a secondary bar, and a nuclear bar \citep{Fathi2007}, which drive gas inwards through successive resonances (Fig.~\ref{fig:gas_dust_panels}). This scenario is supported by the enhanced central molecular gas content, the elevated inner GDR, and the high global $M_{\mathrm{H}_2}/M_{\mathrm{HI}}$ ratio. The high central dust concentration may contribute significantly to efficient HI-to-H$_2$ conversion, fuelling star formation in Region~1 and partially contributing to the depleted atomic gas content in Region~2, as well as the overall enhancement of the H$_2$/HI ratio relative to typical star-forming galaxies.

    \item Dust and molecular gas show a strong spatial correspondence, not only in the centre but also along the spiral arms. The efficient formation of ${\rm H_2}$ catalysed by dust under PDR conditions could help explain this correlation and the central molecular excess. This process could contribute to the radial decline of atomic gas, particularly in Region~2 ($\sim 2$ kpc), and to the GDR more than twice the average value in the central $\sim 1.3$ kpc (GDR $> 200$).
    
    \item The potentially efficient, dust-catalysed conversion of atomic gas may play a significant role in the formation of the distinct nuclear molecular disc ($\sim 1.3$ kpc; Region~1), as reported by \citet{Romeo2015}, which is dynamically separate from the main disc.
\end{enumerate}

In short, the central $\sim 2$ kpc of NGC~6946 are distinct from the rest of the disc, hosting $\sim40\%$ of the stellar mass, the vast majority of the molecular gas and dust, the highest dust temperatures, and the most intense bursts of star formation in its history.  These unique characteristics are  may be the result of a combination of a complex, nested bar-driven dynamical structure and  a potentially efficient, dust-catalysed cycle of atomic-to-molecular gas conversion.

\begin{acknowledgements}
The authors thank the anonymous referee for their constructive comments and valuable suggestions, which helped improve the quality and clarity of the manuscript. SAC (CVU 2054181) thanks SECIHTI for the MSc scholarship. SAC and JF acknowledge the support by PAPIIT-UNAM under grant IN102226. BCS acknowledge the support by PAPIIT-UNAM under grant IN111825. JF, VC, and VT acknowledge funding from the INAF Mini Grant 2024 programme 'DustPedia meets Metal-THINGS: Dust-METAL'. MALL, LEG, MMP, and GV acknowledge grants  PID2024-155875OB-I00 and PID2021-123417OB-I00 funded by MICIU/AEI/10.13039/501100011033/FEDER, EU. MALL  acknowledges grant RYC2020-029354-I funded by MICIU/AEI/10.13039/501100011033 by "ESF Investing in your future", by "ESF+". IZ acknowledges funding from the Deutsche Forschungsgemeinschaft (DFG; German Research Foundation)---project-ID 550945879. LSP acknowledges support from the Research Council of Lithuania (LMTLT) (grant no. P-LU-PAR-26-15). GV acknowledges support from the Comunidad de Madrid through grant PIPF-2024/TEC-34555, and the Atracción de Talento programme 2022-T1/TIC-23797, as well as grant PID2023-146372OB-I00 funded by MICIU/AEI/10.13039/501100011033 and by ERDF, EU. 
\end{acknowledgements}

\bibliographystyle{aa}
\bibliography{references}


\clearpage
\begin{appendix}

\FloatBarrier
\section{Stellar population properties}
\label{AppendixA_SSP}
\begin{figure}[!ht]
    \centering
    \includegraphics[width=.8\linewidth]{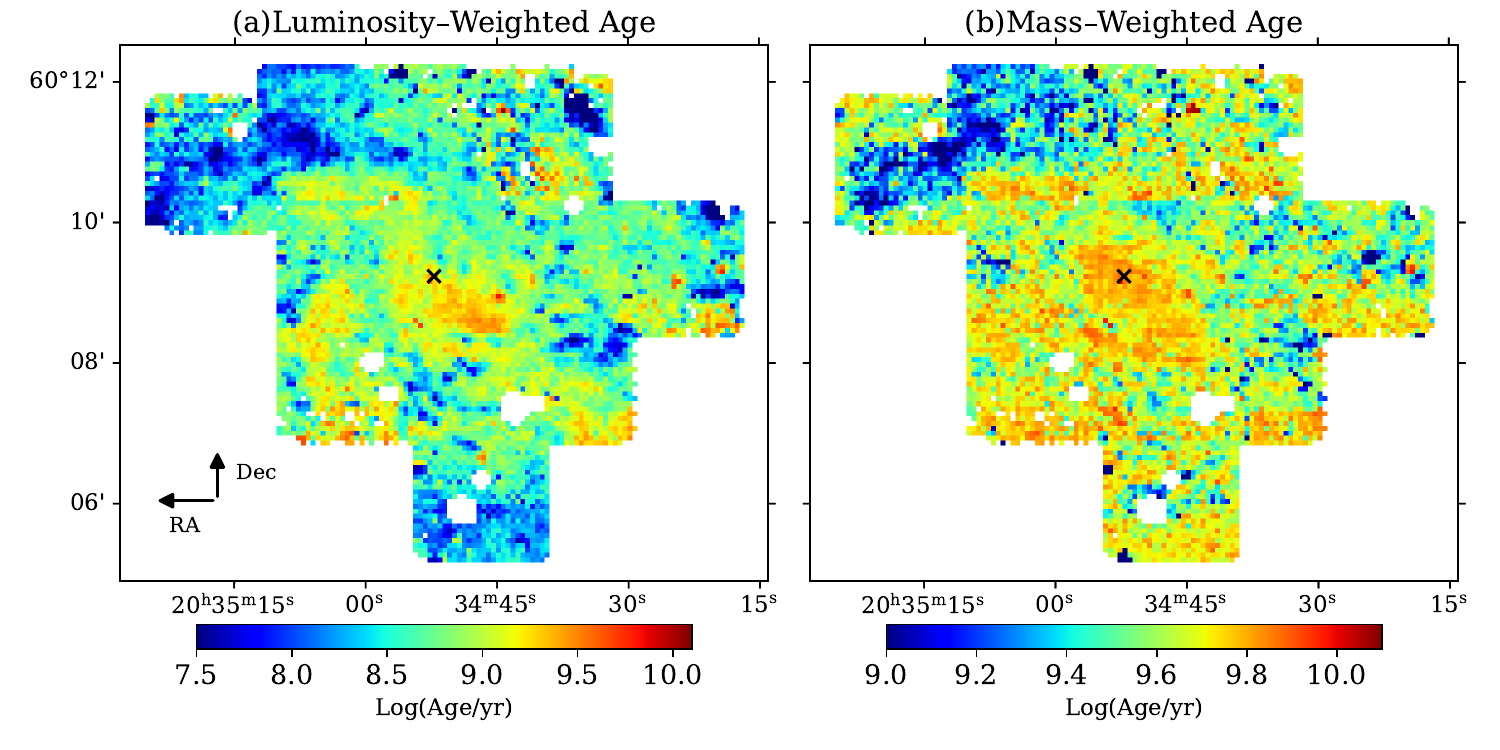}
    \caption{Spatially resolved stellar age maps of NGC~6946. Panel~(a) shows the LW age, while panel~(b) shows the MW age. The galaxy centre is marked with a black $\times$.}
    \label{fig:stellar_ages}
\end{figure}

\begin{figure}[!ht]
    \centering
    \includegraphics[width=0.75\linewidth]{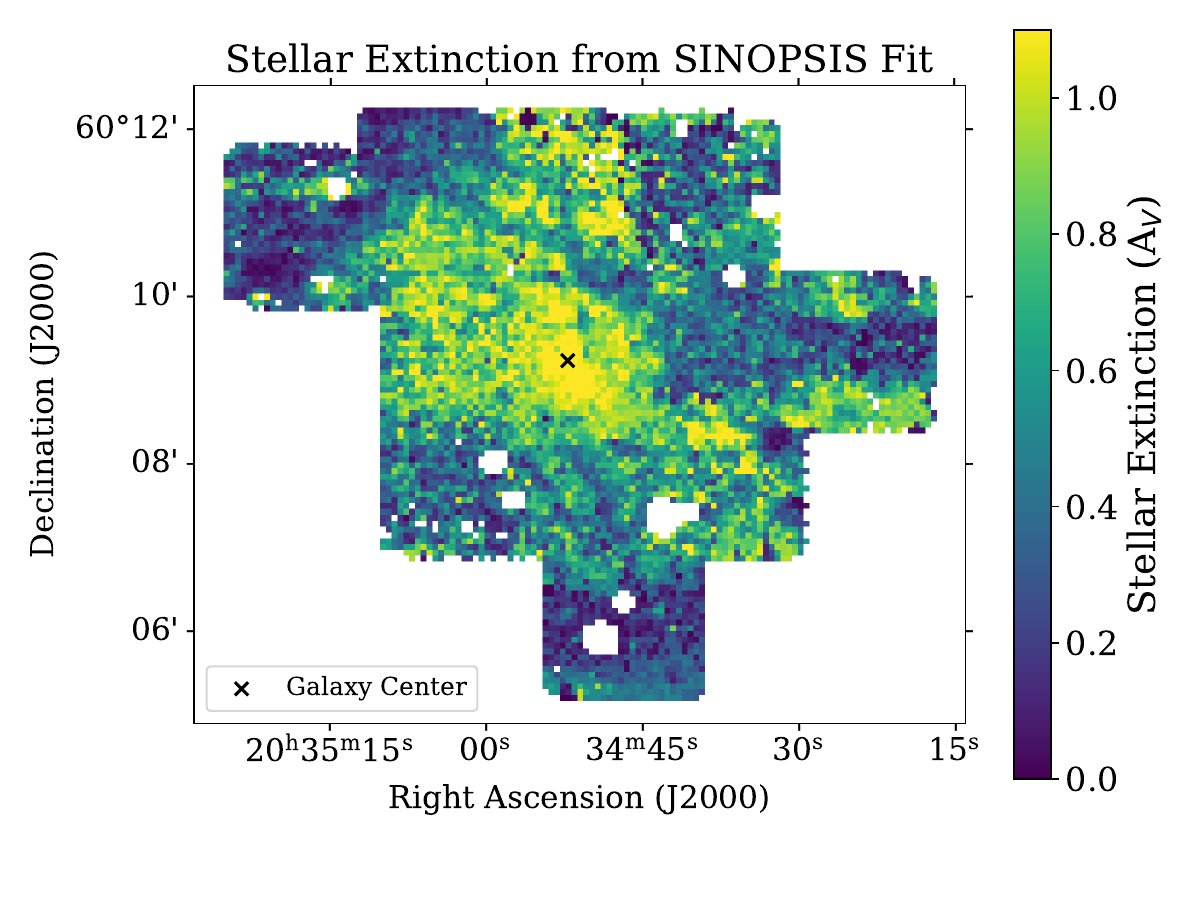}
    \caption{Dust extinction derived from SED fitting with \texttt{SINOPSIS} for the stellar continuum.}
    \label{fig:ext_sinopsis}
\end{figure}

\begin{figure}
    \centering
    \includegraphics[width=0.52\linewidth]{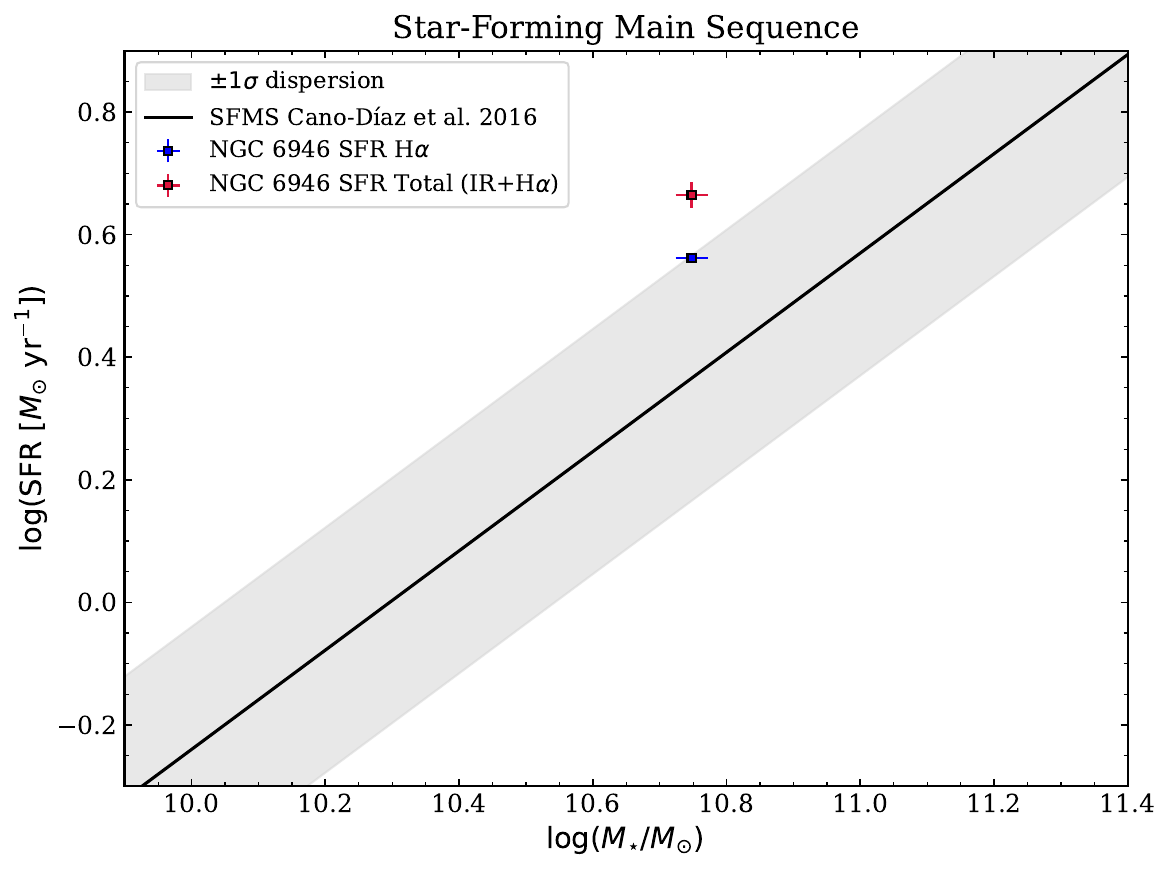}
    \caption{Position of NGC~6946 on the SFMS \citep{Cano2016A}. The blue point corresponds to the recent SFR  from extinction-corrected H$\alpha$ fluxes. The red point corresponds to the total SFR derived from the IR luminosity and the uncorrected H$\alpha$ luminosity. The error bars on the stellar mass were scaled by a factor of 10 for visualisation purposes.}
    \label{fig:sfms}
\end{figure}

From optical spectral fitting, we also derived spatially resolved maps of LW and MW ages (Fig.\ref{fig:stellar_ages}), and continuum extinction for NGC~6946  (Fig.\ref{fig:ext_sinopsis}). The central region shows $\log(\mathrm{LW\mbox{-}age/yr}) \sim 9$, with even younger ages within the inner $\sim0.3$ kpc, coincident with the nuclear stellar bar and consistent with a recent star formation burst \citep{Schinnerer2007}.  The extinction map reveals high central attenuation, reflecting the large dust content that we also find with different methods. In addition, Fig. \ref{fig:sfms} presents the location of NGC~6946 on the SFMS, where it lies above the relation, although bar-driven quenching may already be underway.

\FloatBarrier

\FloatBarrier
\section{Dust properties}
\label{AppendixB_dust}

\begin{figure}[!ht]
    \centering
    \includegraphics[width=0.72\linewidth]{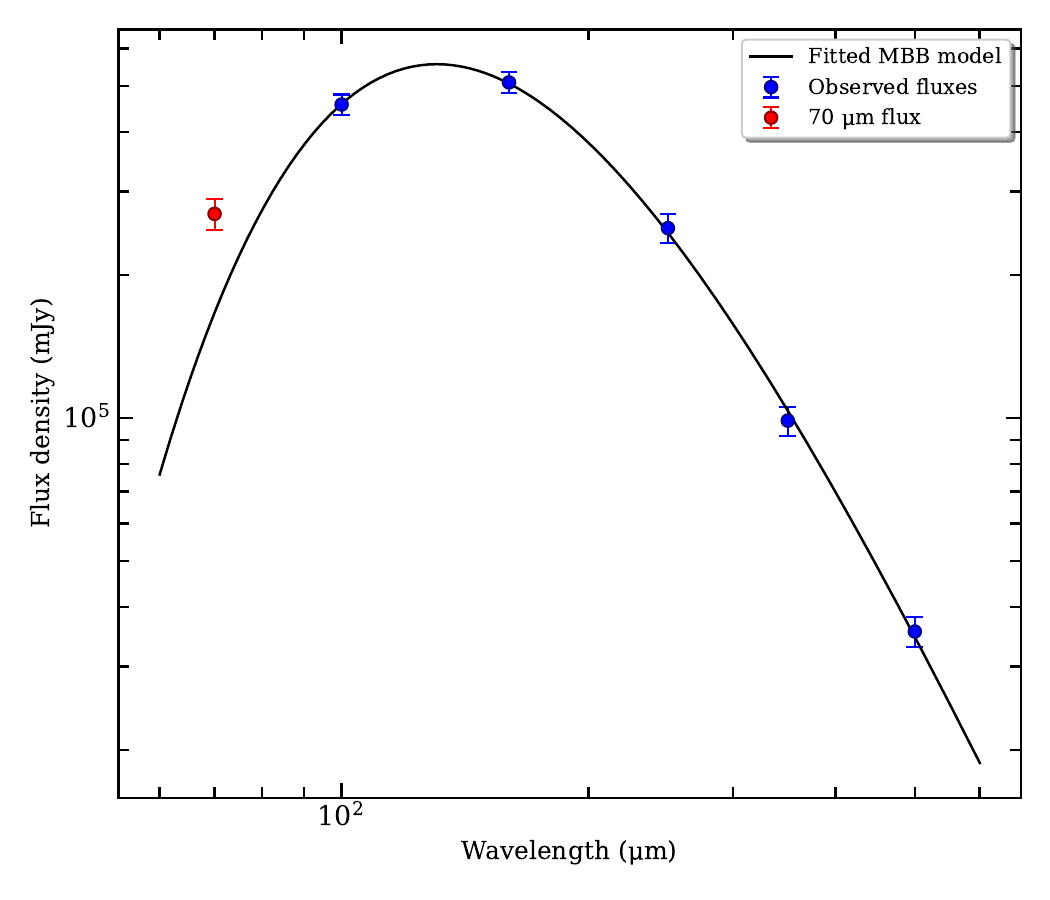}
    \caption{Modified black-body (MBB) fit using the integrated FIR fluxes at 100, 160, 250, 350, and 500~$\mu$m.}
    \label{mbb_total_fit}
\end{figure}

\begin{figure}[!ht]
    \centering
    \includegraphics[width=0.72\linewidth]{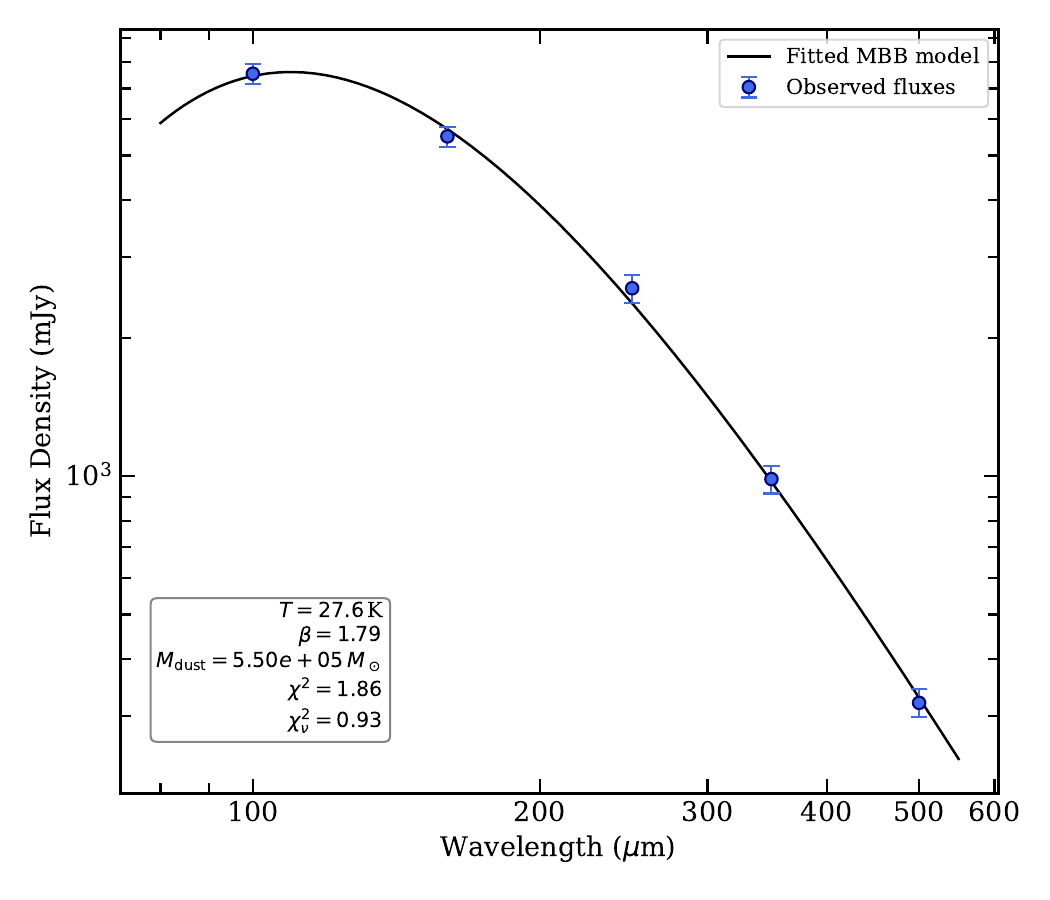}
    \caption{Example of the MBB fit for a central pixel used in the spatially resolved dust analysis.}
    \label{mbb_pixel_fit}
\end{figure}

 Figure~\ref{mbb_total_fit} shows the MBB fit using the total FIR fluxes at 100, 160, 250, 350, and 500~$\mu$m (see Table~\ref{tab:fluxes}). For the spatially resolved analysis, from which we derive the total dust mass, the same procedure was applied on a pixel-by-pixel basis. An example of the fit for a central pixel is shown in Fig.~\ref{mbb_pixel_fit}.
 Table~\ref{tab:dust_properties} summarises the dust properties obtained from both integrated and spatially resolved MBB fits using five \textit{Herschel} bands, as well as a two-band approach (70–100$\mu$m) aimed at recovering dust geometry at higher spatial resolution. The derived dust masses show systematic variations across methods (see Section~\ref{sec:dust} for further discussion).

\begin{table}[!ht]
\centering
\caption{Dust properties derived from MBB fitting.}
\label{tab:dust_properties}
\begin{tabular}{lcc}
\hline
\textbf{Method} & \textbf{M$_d$ (10$^7$ M$_\odot$)} & \textbf{T (K)} \\
\hline
\hline
\multicolumn{3}{l}{\textbf{Global fluxes (GF)}} \\
\quad 5-point fit & $9.52 \pm 1.30$ & $21.85 \pm 1.30$ \\
\quad 2-point (70--100 $\mu$m) & $3.65 \pm 0.76$ & $25.55 \pm 0.76$ \\
\hline
\multicolumn{3}{l}{\textbf{Spatially resolved (SR)}} \\
\quad 5-point fit & $10.27 \pm 0.05$ & $20.53 \pm 0.07$ \\
\quad 2-point (70--100 $\mu$m) & $4.27$ & $24.29$ \\
\hline
\end{tabular}
\end{table}

\begin{figure*}
    \centering
    \includegraphics[width=1\textwidth]{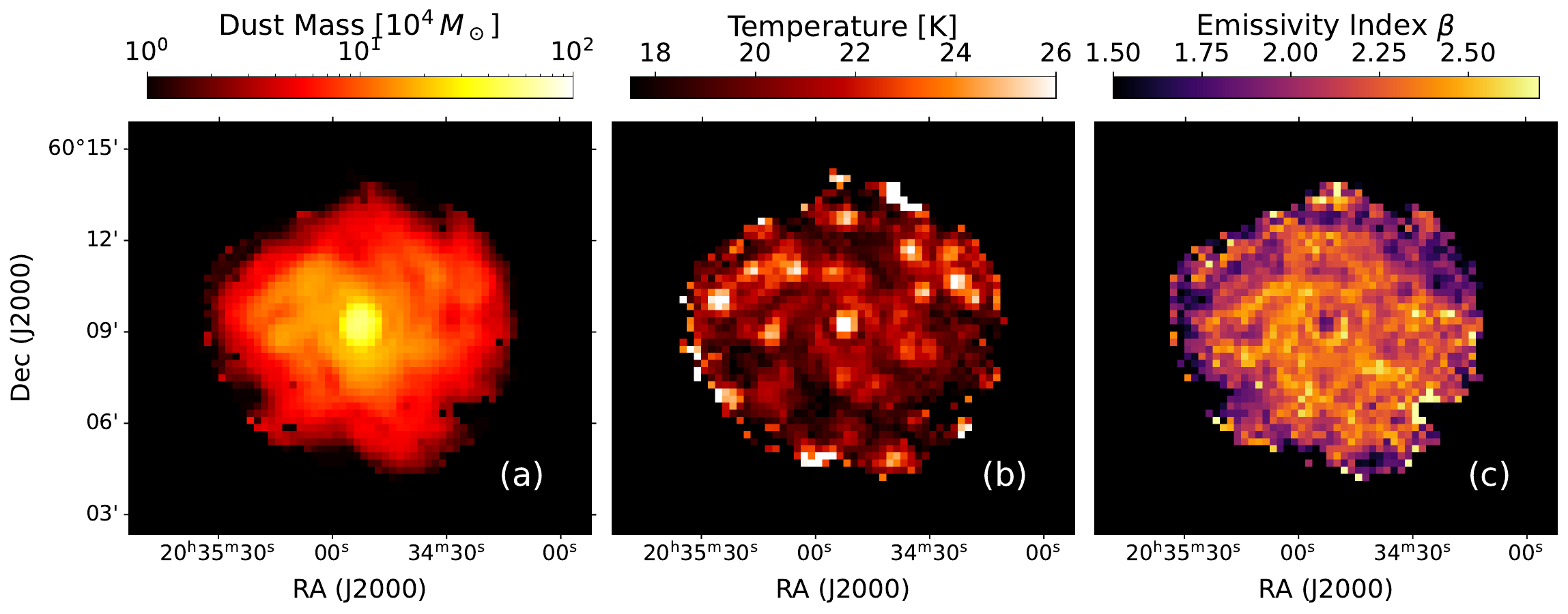}
    \caption{Spatially resolved dust properties (M$_d$, T$_d$, $\beta$, from left to right) of NGC~6946 derived from Herschel flux maps using a MBB model with a 5-point fit for each pixel.}
\label{fig:dust_properties_5pt}
\end{figure*}

Fig.~\ref{fig:dust_properties_5pt} shows the spatially resolved maps of $M_d$, $T_d$, and $\beta$ derived from SED fitting using a MBB model, after smoothing to the SPIRE 500,$\mu$m resolution. We obtain a total dust mass of $M_d = (1.027 \pm 0.144) \times 10^{8}M_{\odot}$ and a mean dust temperature of $T_d = 20.53$K. This value is approximately 9\% higher than that derived from the global integrated flux analysis.

\FloatBarrier
\section{Gas distribution}
\label{AppendixC_gas}

\begin{figure}
    \centering
    \begin{minipage}{0.8\linewidth}
        \centering
        \includegraphics[width=\linewidth]{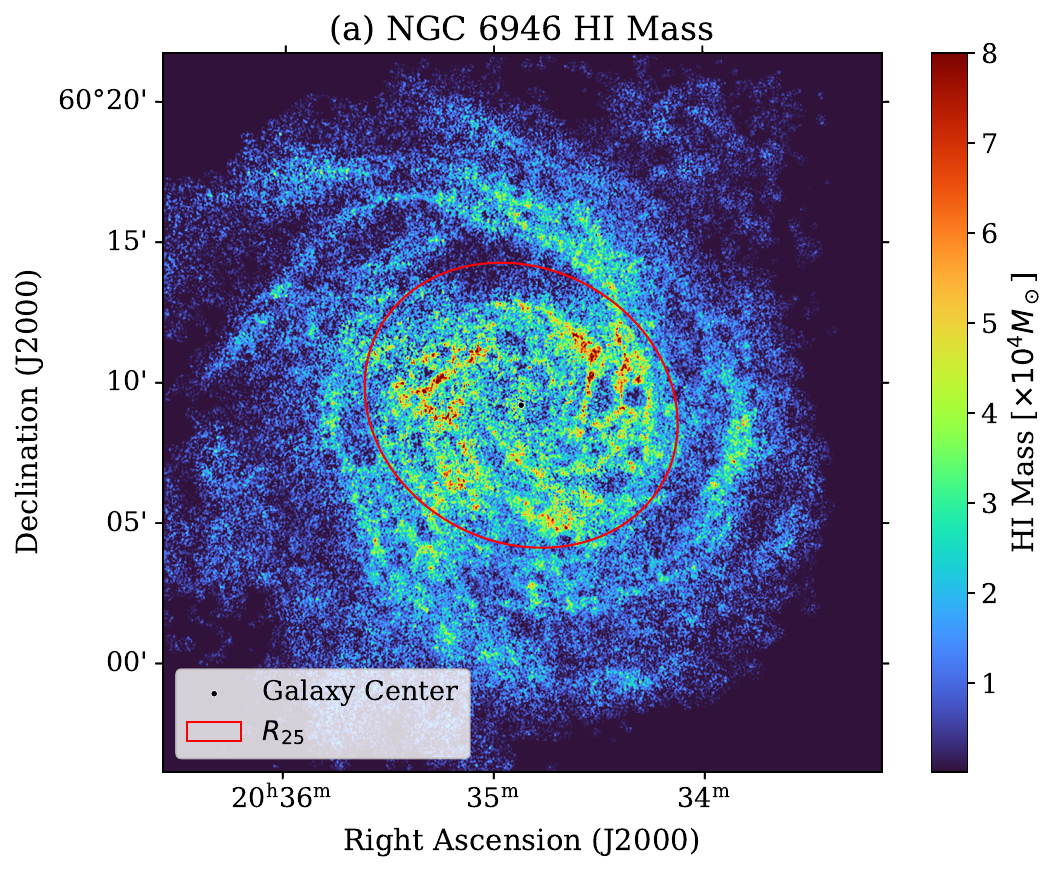}
        
        \vspace{0.2cm}
        
        \includegraphics[width=\linewidth]{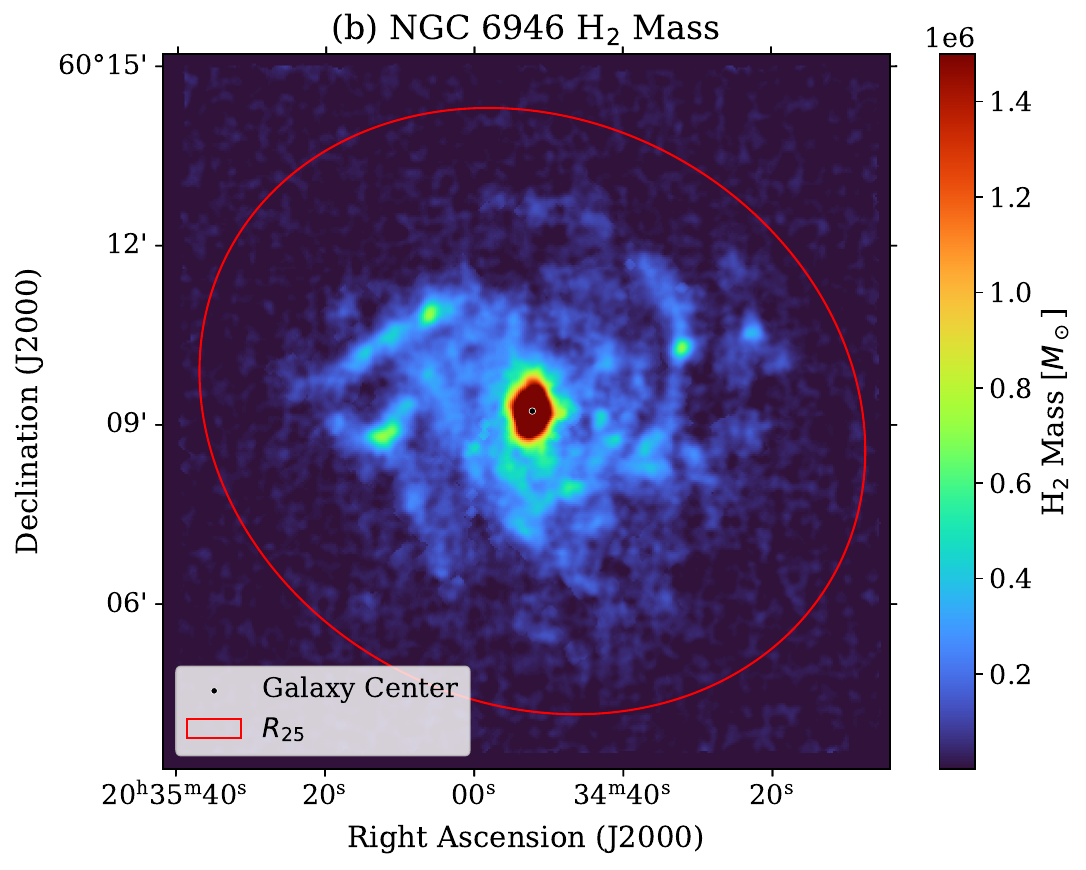}
    \end{minipage}
    \caption{Spatially resolved HI and H$_2$ mass distributions of NGC~6946. \textit{Top panel:} HI extends significantly beyond the stellar disc ($R_{25}$, red ellipse), showing prominent spiral structure and numerous holes attributed to stellar feedback mechanisms. \textit{Bottom panel:} H$_2$ is concentrated within the stellar disc, with a strong central concentration and clear spiral arm features.}
    \label{fig:hi_h2_maps}
\end{figure}

The HI and H$_2$ maps are shown in Fig.~\ref{fig:hi_h2_maps}. We derive total masses of $M_{\mathrm{HI}} = (8.13 \pm 0.71)\times10^9M_{\odot}$ and $M_{\mathrm{H}2} = (7.49 \pm 2.43)\times10^9M{\odot}$. These values indicate an enhanced molecular gas content compared to typical star-forming galaxies, suggesting efficient HI-to-H$_2$ conversion. We attribute this to the combined effects of the bar and dust content (see Section~\ref{sec:discussion_atomictomolecular}).

\end{appendix}

\end{document}